# A NEW H I SURVEY OF ACTIVE GALAXIES

Luis C. Ho[1], Jeremy Darling[2], and Jenny E. Greene[3,4]

## ABSTRACT

We have conducted a new Arecibo survey for H I emission for 113 galaxies with broad-line (type 1) active galactic nuclei (AGNs) out to recession velocities as high as $\sim 35,000$ km s$^{-1}$. The primary aim of the study is to obtain sensitive H I spectra for a well-defined, uniformly selected sample of active galaxies that have estimates of their black hole masses in order to investigate correlations between H I properties and the characteristics of the AGNs. H I emission was detected in 66 out of the 101 (65%) objects with spectra uncorrupted by radio frequency interference, among which 45 (68%) have line profiles with adequate signal-to-noise ratio and sufficiently reliable inclination corrections to yield robust deprojected rotational velocities. This paper presents the basic survey products, including an atlas of H I spectra, measurements of H I flux, line width, profile asymmetry, optical images, optical spectroscopic parameters, as well as a summary of a number of derived properties pertaining to the host galaxies. To enlarge our primary sample, we also assemble all previously published H I measurements of type 1 AGNs for which can can estimate black hole masses, which total an additional 53 objects. The final comprehensive compilation of 154 broad-line active galaxies, by far the largest sample ever studied, forms the basis of our companion paper, which uses the H I database to explore a number of properties of the AGN host galaxies.

*Subject headings:* galaxies: active — galaxies: bulges — galaxies: ISM — galaxies: kinematics and dynamics — galaxies: nuclei — galaxies: Seyfert

## 1. INTRODUCTION

Central black holes (BHs) with masses ranging from $\sim 10^6$ to a few $\times 10^9$ $M_\odot$ are an integral component of most, perhaps all, galaxies with a bulge component (Kormendy 2004), and although rarer, at least some late-type galaxies host nuclear BHs with masses as low as $\sim 10^5$ $M_\odot$ (Filippenko & Ho 2003; Barth et al. 2004; Greene & Ho 2007a, 2007b). It is now widely believed that BHs play an important role in the life cycle of galaxies (see reviews in Ho 2004). To date, most of the observational effort to investigate the relationship between BHs and their host galaxies have focused on the stellar component of the hosts, especially the velocity dispersion and luminosity of the bulge, which empirically seem most closely coupled to the BH mass. Although the gas content of inactive galaxies has been extensively studied (e.g., Haynes & Giovanelli 1984; Knapp et al. 1985; Roberts et al. 1991; Bregman et al. 1992; Morganti et al. 2006), comparatively little attention has been devoted to characterizing the interstellar medium of active galaxies or of systems with knowledge of their BH mass or accretion rate.

The gaseous medium of the host galaxy, especially the cold phase as traced in neutral atomic or molecular hydrogen, offers a number of diagnostics inaccessible by any other means. Since cold gas constitutes the very raw material out of which both the stars form and the BH grows, the cold gas content of the host galaxy is one of the most fundamental quantities that can be measured in the effort to understand the coevolution of BHs and galaxies. At the most rudimentary level, we might naively expect the gas content of the host to be correlated with the BH accretion rate or the luminosity of its active galactic nucleus (AGN). Likewise, the gas content should reflect the particular evolutionary stage of the host galaxy. Many current models (e.g., Granato et al. 2004; Springel et al. 2005) invoke AGN feedback as a key ingredient for galaxy formation and for coupling the BH to its host. Depending on the violence with which the accretion energy is injected into the host and the evolutionary state of the system, AGN feedback can wreck havoc on the interstellar medium of the host. For example, recent H I absorption observations of radio-loud AGNs detect substantial quantities of high-velocity outflowing neutral gas, presumably in the midst of being expelled from the host galaxy by the radio jet (Morganti et al. 2007). Performing a careful, systematic census of the cold gas content of AGN hosts will provide much needed empirical guidance for AGN feedback models. Apart from the sheer gas mass, H I and CO observations, even when conducted in spatially unresolved mode, can provide other useful probes of the physical properties of the host, and of its circumgalactic environment (e.g., Ho 2007a, 2007b). For example, the width of the integrated line profile, if it is sufficiently regular, gives an estimate of the rotation velocity of the disk, and hence an additional handle on the gravitational potential of the system. Combining the line width with the Tully-Fisher (1977) relation, we can infer immediately the total luminosity of the host, independent of any contamination from the AGN. The degree of symmetry of the line profile furnishes useful, if crude, information on the spatial distribution of gas within and around the host, as well as an effective probe of possible dynamic disturbances due to neighboring galaxies.

The primary goal of this study is to quantify the H I content of a large, well-defined sample of active galaxies with uniformly measured BH masses and optical properties, spanning a wide range in AGN properties. Despite the obvious importance of understanding the cold gas component of AGN host galaxies, there has been relatively little modern work conducted with this explicit goal in mind. Although there have been a number of H I surveys of AGNs, most of them have focused on relatively low-luminosity Seyfert nuclei (Allen et al. 1971; Heckman et al. 1978; Bieging & Biermann 1983; Mirabel & Wilson 1984; Hutchings 1989; Greene et al. 2004) and radio-emitting elliptical galaxies (Dressel et al. 1982; Jenkins 1983), with only limited attention devoted to higher luminosity quasars (Condon et al. 1985; Hutchings et al. 1987; Lim & Ho 1999). This is in part due to sensitivity limitations (quasars are more distant), but also due to the poor baselines of pre-upgrade Arecibo[5] spectra. With the new Gregorian optics, $L$-band receiver, and modern backend at Arecibo, the time is ripe to revisit the problem in a concerted fashion. In light of the scientific issues outlined above, the motivation has never been stronger. We are particularly keen to use the H I line width as a kinematic tracer of the host galaxy potential. Since the rotation velocity of the disk is correlated with the stellar velocity dispersion of the bulge (see Ho 2007a, and references therein), the H I line width can be used as a new variable to investigate the correlation between BH mass and galaxy potential. We are additionally interested in using the H I spectra to obtain dynamical masses for the host galaxies, to use the line shape to probe the nearby environment and dynamical state of the hosts, and to evaluate possible correlations between H I content and AGN properties. These issues are investigated in a companion paper (Ho et al. 2008).

[1] The Observatories of the Carnegie Institution of Washington, 813 Santa Barbara St., Pasadena, CA 91101.
[2] Center for Astrophysics and Space Astronomy, Department of Astrophysical and Planetary Sciences, University of Colorado, 389 UCB, Boulder, CO 80309-0389.
[3] Princeton Observatory, Peyton Hall, Princeton University, Princeton, NJ 08544-1001.
[4] Hubble Fellow and Carnegie-Princeton Fellow.

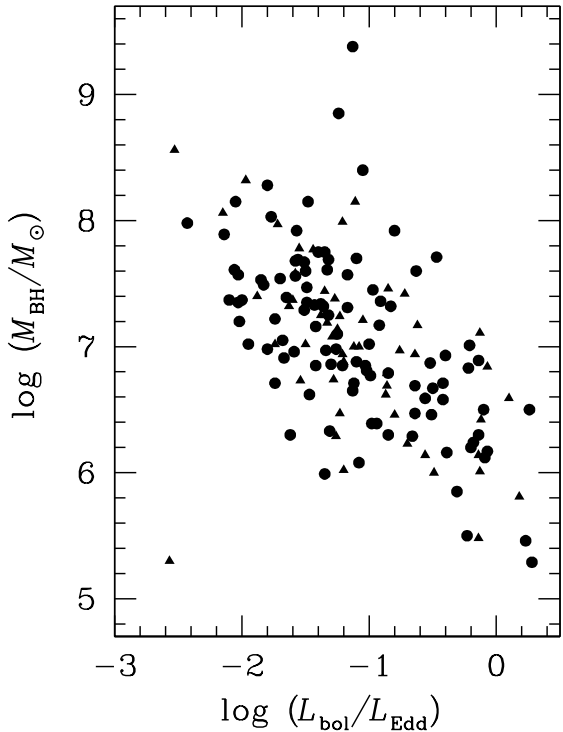

FIG. 1.— The distribution of BH masses and Eddington ratios for the sample included in this study. The 101 newly surveyed objects for which H I observations were successfully obtained are plotted as circles, while the sample of 53 sources taken from the literature are marked as triangles.

## 2. OBSERVATIONS AND DATA REDUCTION

### 2.1. Sample

Our sample of AGNs was chosen with one overriding scientific motivation in mind: the availability of a reliable BH mass estimate. As we rely on the virial mass method to estimate BH masses (Kaspi et al. 2000; Greene & Ho 2005b; Peterson 2007), this limits our targets to type 1 AGNs. Sensitivity considerations with the current Arecibo system imposes a practical redshift limit of $z \lesssim 0.1$. Apart from these two factors, and the visibility restrictions of Arecibo ($0° \lesssim \delta \lesssim 37°$), the targets were selected largely randomly to fill the available schedule blocks of the telescope. The 113 newly observed objects, whose basic properties are summarized in Table 1, contains two subsamples. The first comprises 98 type 1 AGNs from the Fourth Data Release of the Sloan Digital Sky Survey (SDSS; Adelman-McCarthy et al. 2006), which form part of an on-going study of low-redshift AGNs by Greene (2006; see also Greene & Ho 2004, 2005b, 2006a, 2006b, 2007a, 2007b). Although the SDSS objects strictly do not form a complete or unbiased sample, they are representative of low-redshift broad-line AGNs of moderate to high luminosities. With $M_g \approx -18.8$ to $-23.1$ mag, only $\sim 3-4$ objects satisfy the conventional luminosity threshold of quasars[6], but most are very prominent Seyfert 1 nuclei. Twenty-eight of the objects have broad H$\alpha$ profiles with full-width at half maximum (FWHM) less than 2000 km s$^{-1}$, and thus meet the formal line width criterion of narrow-line Seyfert 1 galaxies (e.g., Osterbrock & Pogge 1985). The second subsample, in total 15 objects, were primarily chosen because they have been studied with reverberation mapping (Kaspi et al. 2000; Peterson et al. 2004); we deem these to be high-priority objects because they have better-determined BH masses. This subsample includes seven Palomar-Green (PG) sources (Schmidt & Green 1983), among them five luminous enough to qualify as bona fide quasars, and two satisfying the line width criterion of narrow-line Seyfert 1 galaxies (PG 0003+199 and PG 1211+143).

To augment the sample size and to increase its dynamic range in terms of BH mass and AGN luminosity, we performed a comprehensive search of the literature to compile all previously published H I measurements of type 1 AGNs that have sufficient optical data to allow estimation of BH masses. The results of this exercise yielded a sizable number of additional objects (53), the details of which are documented in the Appendix. Our final sample, now totaling 166 and by far the largest ever studied, covers a wide range of BH masses, from $M_{\rm BH} \approx 10^5$ to $10^9$ $M_\odot$, and a significant spread in Eddington ratios, from $\log L_{\rm bol}/L_{\rm Edd} \approx -2.7$ to 0.3 (Fig. 1), where $L_{\rm Edd} \equiv 1.26 \times 10^{38} \left(M_{\rm BH}/M_\odot\right)$ ergs s$^{-1}$. Although the sample definitely contains predominantly low-luminosity AGNs, it covers at least 4 orders of magnitude in nuclear luminosity (Fig. 2a), from $L_{\rm H\alpha} \approx 10^{40}$ to $10^{44}$ ergs s$^{-1}$ (excluding the ultra-low-luminosity object NGC 4395 at $L_{\rm H\alpha} \approx 10^{38}$ ergs s$^{-1}$), which in more familiar units corresponds to $B$-band absolute magnitudes of $M_B \approx -15.5$ to $-24.75$ mag (Fig. 2b).

### 2.2. Arecibo Observations

We observed the 21 cm spin-flip transition of neutral hydrogen (H I) in our sample at the Arecibo radio telescope from

---

[5] The Arecibo Observatory is part of the National Astronomy and Ionosphere Center, which is operated by Cornell University under a cooperative agreement with the National Science Foundation.

[6] The canonical luminosity threshold of quasars, $M_B = -23.0$ mag (Schmidt & Green 1983), translates to $M_B = -22.1$ mag in our distance scale, which assumes $H_0 = 70$ km s$^{-1}$ Mpc$^{-1}$, $\Omega_m = 0.3$, and $\Omega_\Lambda = 0.7$. For a power-law AGN spectrum of the form $f_\lambda \propto \lambda^{-1.56}$ (Vanden Berk et al. 2001), this threshold is $M_g \approx -22.3$ mag.

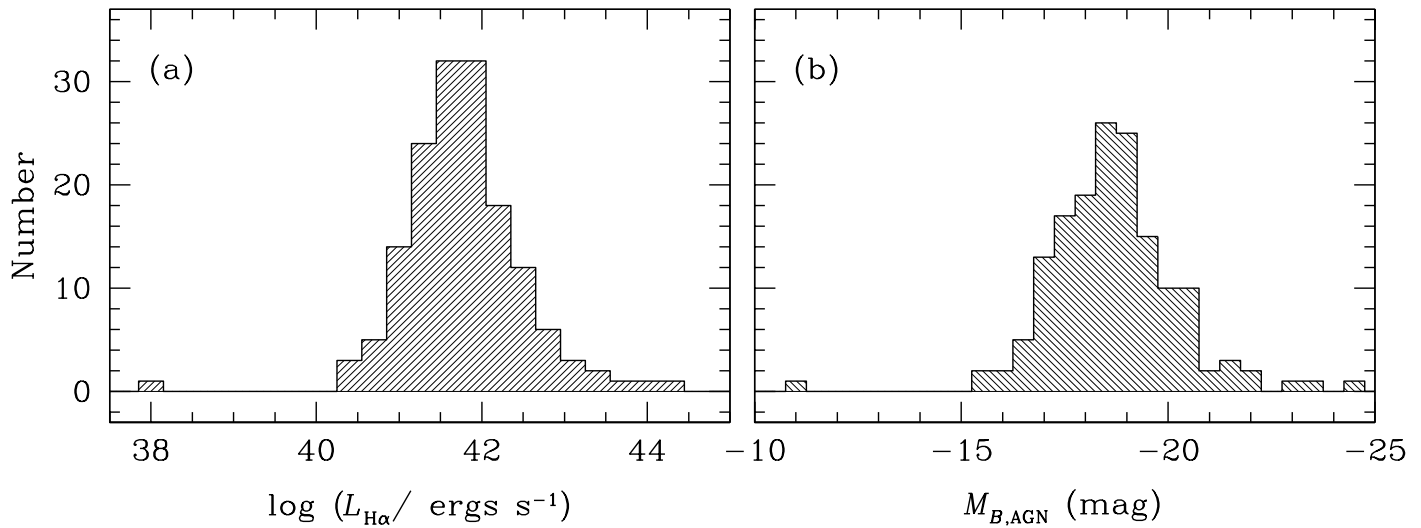

FIG. 2.— The distribution of (a) Hα luminosity and (b) B-band absolute magnitude of the AGN component for the sample objects with H I data. The Hβ luminosities of the literature sample (Table 6) were converted to Hα assuming Hα/Hβ = 3.5, as empirically determined by Greene & Ho (2005b). To convert between Hα luminosity and B-band absolute magnitude, we employ the correlation between Hα and 5100 Å continuum luminosity of Greene & Ho (2005b), and then assume a continuum spectrum of the form $f_\lambda \propto \lambda^{-1.56}$ (Vanden Berk et al. 2001) to extrapolate to 4400 Å.

November 2005 through April 2007. The H I observations were conducted in four independently tracked 25 MHz bands centered on redshifted H I and OH (1420.405751786, 1612.2310, 1667.3590, and 1720.5300 MHz, respectively). Observations consisted of 5-minute position-switched scans, with a calibration diode fired after each position-switched pair and spectral records recorded every 6 seconds. Typically sources were observed for 1–2 hours. The autocorrelation spectrometer used 1024 channels and 9-level sampling in two (subsequently averaged) polarizations. Rest-frame velocity resolutions ranged from 5.15 km s$^{-1}$ ($z = 0$) to 5.72 km s$^{-1}$ ($z = 0.11$), but most spectra were Hanning smoothed, reducing the velocity resolution roughly by a factor of 2.

Records were individually calibrated and bandpasses flattened using the calibration diode and the corresponding off-source records. Records and polarizations were subsequently averaged, and a low-order polynomial baseline was fit and subtracted. Systematic flux calibration errors in these data are of order 10%. All data reduction was performed in AIPS++[7]. Some spectra showed standing waves due to resonances within the telescope superstructure or strong continuum sources ($\gtrsim$ 300 mJy) falling in the beam (either coincidentally or due to strong radio emission from the target galaxy itself). The expected H I line widths are similar to the size of the standing wave features ($\sim 1$ MHz), so the detectability of lines was severely impaired in a few cases.

The observed bands were generally interference-free in the vicinity of the observed lines, requiring little or no flagging, but some redshift ranges, most prominently $z \simeq 0.066$–0.069 and $z \simeq 0.051$–0.054, were unobservable due to radio frequency interference (RFI). Hence, the redshift distribution of the sample has gaps. H I lines were detected in 66 galaxies in the sample, 35 galaxies were significant nondetections, and 12 galaxies are indeterminate due to standing waves in the bandpass or RFI.

The spectra for the detected sources are plotted in Figure 3, accompanied by their optical images. No 18 cm OH lines were detected in the sample (many lines were unobservable due to RFI).

The H I properties of the sample are summarized in Table 2. For each detected source, we list the systemic velocity of the line in the barycentric frame ($v_{sys}$), defined to be the midpoint (mean) of the velocities corresponding to the 20% point of the two peaks in the H I profile, and the line width $W_{20}$, the difference between these two velocities. The actual high and low velocities are obtained from an interpolation between the two data points bracketing 20% of peak flux. For a typical root-mean-square noise level of $\sim 0.3$ mJy, we estimate that the uncertainty in the systemic velocity is $\sigma(v_{sys}) \approx 3.4$ km s$^{-1}$; the uncertainty in the line width is $\sigma(W_{20}) = 2\sigma(v_{sys}) \approx 6.8$ km s$^{-1}$. In practice, these formal values underestimate the true errors for spectra affected by RFI or poor bandpasses, or in instances when the line profile is not clearly double-peaked. Profiles that are single-peaked and/or highly asymmetric are noted in Table 2.

To convert the raw line widths to $v_m$, the maximum rotational velocity, four corrections must be applied to $W_{20}$: (1) instrumental resolution, which we assume to be $W_{inst} = 10$ or 5 km s$^{-1}$, depending on whether the spectrum was Hanning smoothed or not, and that it can be removed by linear subtraction; (2) redshift, which stretches the line width by a factor $(1+z)$; (3) turbulent broadening, which for simplicity we assume to be $W_{turb} = 22$ km s$^{-1}$ for $W_{20}$ and can be subtracted linearly (Verheijen & Sancisi 2001); and (4) inclination angle. We assume that the inclination of the H I-emitting disk to the line-of-sight can be approximated by the photometric inclination angle of the optical disk, $i$ (see §2.3). The final maximum rotational velocity is then

[7] The AIPS++ (Astronomical Information Processing System) is freely available for use under the Gnu Public License. Further information may be obtained from http://aips2.nrao.edu.

TABLE 1: THE SAMPLE

| Source Name | Alternate Name | $C$ | Type | $q$ | $i$ (°) | $D_{25}$ (″) | $z$ | $D_L$ (Mpc) | $M_g$ (mag) | $g-r$ (mag) | $\sigma_*\pm$error (km s$^{-1}$) | FWHM$_{\mathrm{H}\alpha}$ (km s$^{-1}$) | log $L_{\mathrm{H}\alpha}$ ($L_\odot$) | log $M_{\mathrm{BH}}$ ($M_\odot$) | log $L_{\mathrm{bol}}$ ($L_{\mathrm{Edd}}$) |
|---|---|---|---|---|---|---|---|---|---|---|---|---|---|---|---|
| (1) | (2) | (3) | (4) | (5) | (6) | (7) | (8) | (9) | (10) | (11) | (12) | (13) | (14) | (15) | (16) |
| SDSS J000805.62+145023.3 | ⋯ | 0.37 | SBb | 0.45 | 69.5 | 17.8 | 0.0454 | 201.6 | −20.93 | 0.18 | 140±27 | 7211 | 41.17 | 7.61 | −2.06 |
| SDSS J003646.45+145936.9 | ⋯ | 0.44 | S0/Sp | 0.85 | 33.5 | 6.9 | 0.0892 | 408.8 | −20.81 | −0.44 | (149±15) | 1220 | 41.42 | 6.16 | −0.39 |
| SDSS J004055.88+153349.0 | ⋯ | 0.43 | S0/Sp | 0.89 | 28.9 | 6.9 | 0.0988 | 455.7 | −21.49 | 0.11 | 124±21 | 532 | 41.19 | 5.29 | 0.28 |
| SDSS J004719.39+144212.6 | Mrk 1146 | 0.44 | SBbc | 0.63 | 54.8 | 25.9 | 0.0393 | 173.7 | −21.99 | −0.17 | 180±42 | 2185 | 41.62 | 6.79 | −0.85 |
| SDSS J010712.03+140844.9 | ⋯ | 0.37 | E/S0 | 0.90 | 27.7: | 4.6 | 0.0768 | 349.0 | −19.87 | −0.01 | 128±26 | 880 | 42.57 | 6.50 | 0.26 |
| SDSS J015046.68+132359.9 | | 0.41 | E/S0 | 0.66 | 52.2: | 3.8 | 0.0941 | 432.8 | −20.24 | 0.30 | 90±33 | 1085 | 41.76 | 6.24 | −0.18 |
| SDSS J075245.60+261735.7 | ⋯ | 0.38 | E/S0 | 0.96 | 17.8: | 5.5 | 0.0822 | 374.6 | −21.35 | −0.17 | (94±9) | 1139 | 42.15 | 6.50 | −0.10 |
| SDSS J080243.39+310403.3 | | 0.34 | E/S0 | 0.84 | 34.2 | 10.6 | 0.0409 | 181.0 | −20.80 | 0.47 | 151±17[a] | 5191 | 41.84 | 7.69 | −1.56 |
| SDSS J080538.66+261005.4 | IC 492 | 0.51 | SBcd | ⋯ | 25 | 62.8 | 0.0169 | 73.6 | −20.63 | 0.20 | 100±14 | 3108 | 40.16 | 6.30 | −1.62 |
| SDSS J080546.97+260532.9 | ⋯ | 0.33 | S0 | ⋯ | 20: | 6.7 | 0.0738 | 334.4 | −20.98 | 0.25 | 91±28 | 2395 | 41.43 | 6.77 | −0.99 |
| SDSS J081700.40+343556.3 | | 0.28 | E/S0 | 0.83 | 36.3: | 9.8 | 0.0617 | 277.4 | −20.63 | 0.86 | 119±21 | 4302 | 42.01 | 7.61 | −1.33 |
| SDSS J081835.73+285022.4 | ⋯ | 0.37 | S0/Sp | 0.82 | 36.5: | 6.9 | 0.0772 | 350.9 | −20.77 | 0.35 | 135±28 | 1646 | 41.49 | 6.47 | −0.64 |
| SDSS J082048.30+282217.8 | | 0.42 | Sp(d) | 0.71 | 47.7 | 4.9 | 0.0932 | 428.4 | −20.22 | 0.68 | 187±21 | 2560 | 41.22 | 6.71 | −1.12 |
| SDSS J082320.68+074620.2 | | 0.56 | SBc | 0.96 | 16.6 | 12.6 | 0.0646 | 290.8 | −21.39 | 0.66 | 103±18 | 10852 | 41.17 | 7.98 | −2.43 |
| SDSS J083045.36+340532.1 | | 0.31 | SBa | 0.47 | 67.4 | 11.3 | 0.0623 | 279.9 | −21.10 | 0.31 | 210±46 | 1772 | 42.10 | 6.87 | −0.52 |
| SDSS J083107.63+052105.9 | | 0.30 | E/S0 | 0.87 | 30.8 | 7.6 | 0.0635 | 285.7 | −20.09 | 0.97 | 223±30 | 3811 | 41.39 | 7.16 | −1.42 |
| SDSS J083747.89+053644.8 | | 0.34 | E/S0 | 0.69 | 49.2 | 5.9 | 0.0991 | 457.0 | −20.57 | 0.82 | 127±32 | 5020 | 41.87 | 7.67 | −1.51 |
| SDSS J084025.51+033301.7 | | 0.35 | E/S0 | 0.85 | 33.1: | 8.1 | 0.0605 | 271.6 | −20.81 | 0.20 | 164±32[b] | 1544 | 42.03 | 6.71 | −0.42 |
| SDSS J084556.67+340936.2 | | 0.38 | S0/Sp | 0.90 | 27.8 | 17.6 | 0.0655 | 295.1 | −21.49 | 0.60 | 102±27 | 1498 | 41.63 | 6.46 | −0.51 |
| SDSS J090028.53+060842.5 | ⋯ | 0.38 | S0/Sp | 0.90 | 27.1: | 5.6 | 0.0773 | 351.1 | −19.94 | 0.30 | 125±21 | 1784 | 41.05 | 6.30 | −0.85 |
| SDSS J091222.31+065829.0 | ⋯ | 0.42 | S0/Sp | 0.91 | 25.9: | 3.8 | 0.0781 | 355.3 | −19.53 | 0.45 | 273±58 | 6960 | 41.16 | 7.57 | −2.03 |
| SDSS J093240.55+023332.6 | | 0.40 | S0/Sp | 0.79 | 40.2 | 9.1 | 0.0568 | 254.2 | −20.79 | 0.66 | 102±17 | 3688 | 41.61 | 7.25 | −1.32 |
| SDSS J093259.60+040506.0 | | 0.41 | S0/Sp | 0.89 | 27.9 | 10.6 | 0.0590 | 264.7 | −21.25 | 0.54 | 70.5±6.7[a] | 4314 | 41.53 | 7.35 | −1.49 |
| SDSS J093812.26+074340.0 | | 0.34 | E/S0 | 0.79 | 49.9 | 16.0 | 0.0219 | 95.5 | −19.97 | 0.72 | 119±8.0[a] | 3365 | 41.03 | 6.85 | −1.42 |
| SDSS J093917.26+363343.9 | MCG +06-21-068 | 0.31 | S0 | 0.68 | 50.3 | 29.3 | 0.0201 | 87.3 | −19.95 | 0.69 | 117±20 | 3146 | 40.71 | 6.62 | −1.47 |
| SDSS J094529.37+093610.4 | MS 0942.8+0950 | 0.32 | SB0 | 0.67 | 50.8 | 18.7 | 0.0133 | 57.6 | −18.82 | 0.63 | 118±20 | 1933 | 40.53 | 6.08 | −1.08 |
| SDSS J095955.85+130237.8 | NGC 3080 | 0.34 | Sa | 0.96 | 17.1: | 47.7 | 0.0354 | 156.1 | −21.69 | 0.45 | 100±54[b] | 1796 | 41.75 | 6.69 | −0.64 |
| SDSS J100155.79+055413.3 | | 0.35 | Sa | ⋯ | 80.0: | ⋯ | 0.1034 | 478.5 | −20.77 | 0.69 | 131±21 | 5254 | 41.81 | 7.68 | −1.58 |
| SDSS J102148.90+030732.2 | | 0.42 | Sa | 0.77 | 41.8 | 10.0 | 0.0618 | 277.8 | −21.27 | 0.53 | (452±45) | 2556 | 41.48 | 6.85 | −1.03 |
| SDSS J102402.60+062943.9 | ⋯ | 0.43 | Sa | 0.93 | 22.8 | 12.5 | 0.0440 | 195.2 | −20.69 | 0.21 | 68±23 | 1561 | 41.26 | 6.29 | −0.66 |
| SDSS J102925.73+140823.2 | | 0.27 | S0 | 0.63 | 54.8 | 14.9 | 0.0608 | 273.0 | −21.61 | 0.61 | 181±25 | 4529 | 41.67 | 7.47 | −1.49 |
| SDSS J104326.47+110524.3 | | 0.40 | S0/Sp | 0.75 | 44.0: | 8.9 | 0.0476 | 211.5 | −20.35 | 0.03 | (262±26) | 4619 | 42.15 | 7.75 | −1.35 |
| SDSS J104913.79+044039.9 | | 0.38 | S0/Sp | 0.67 | 51.3 | 6.6 | 0.0884 | 404.7 | −20.57 | 0.59 | 212±36 | 3272 | 41.52 | 7.10 | −1.25 |
| SDSS J105115.42+054824.6 | ⋯ | 0.40 | Sa | 0.27 | 90.0 | 12.4 | 0.0695 | 313.9 | −20.06 | 0.47 | 173±25 | 3105 | 41.46 | 7.02 | −1.22 |
| SDSS J110538.99+020257.4 | | 0.40 | S0/Sp | 0.65 | 53.0 | 10.7 | 0.1066 | 494.4 | −22.60 | 0.19 | 226±50 | 3830 | 42.83 | 7.96 | −0.97 |
| SDSS J110640.20+051905.6 | | 0.40 | SBb | 0.79 | 39.9 | 13.2 | 0.0913 | 419.1 | −21.99 | 0.60 | 140±15[a] | 5770 | 42.10 | 7.92 | −1.57 |
| SDSS J110654.46+061213.1 | ⋯ | 0.29 | E | 0.85 | 33.5: | 11.4 | 0.0435 | 192.6 | −19.83 | 0.37 | 93±29 | 6260 | 40.96 | 7.37 | −2.00 |
| SDSS J111031.61+022043.2 | ⋯ | 0.41 | E/S0 | 0.81 | 37.6: | 4.6 | 0.0795 | 361.8 | −20.01 | 0.29 | 70±15 | 1016 | 41.16 | 5.85 | −0.31 |
| SDSS J111045.96+113641.6 | | 0.40 | SBb | 0.80 | 39.2 | 25.8 | 0.0423 | 187.4 | −21.48 | 0.56 | 110±18 | 3102 | 41.18 | 6.86 | −1.30 |
| SDSS J111237.43+120729.5 | | 0.35 | Sa | 0.48 | 66.6 | 20.2 | 0.0467 | 207.7 | −20.72 | 0.76 | 189±26 | 4651 | 40.74 | 6.98 | −1.80 |
| SDSS J111639.15+040427.6 | ⋯ | 0.32 | E | 0.81 | 37.5: | 7.5 | 0.0745 | 337.9 | −20.32 | 0.10 | 67±22 | 2483 | 41.45 | 6.81 | −1.02 |
| SDSS J112328.11+052823.2 | | 0.42 | E/S0 | 0.85 | 33.5 | 4.7 | 0.1013 | 468.3 | −20.85 | 0.29 | 232±80[b] | 1516 | 42.00 | 6.67 | −0.40 |
| SDSS J112455.86+080615.3 | | 0.44 | SBb | 0.77 | 42.3 | 9.7 | 0.0701 | 316.7 | −21.68 | 0.50 | 118±29 | 3970 | 40.96 | 6.96 | −1.59 |
| SDSS J112555.85+073107.8 | | 0.36 | SBa | 0.90 | 26.5 | 8.3 | 0.0749 | 337.7 | −21.04 | 0.59 | 121±17 | 4915 | 41.39 | 7.39 | −1.65 |
| SDSS J112813.02+102308.3 | ⋯ | 0.42 | S0/Sp | 0.98 | 12.2: | 6.9 | 0.0504 | 224.7 | −20.80 | 0.05 | 113±113 | 1079 | 41.87 | 6.30 | −0.14 |
| SDSS J112904.01+061140.0 | | 0.32 | E/S0 | 0.94 | 20.1 | 6.3 | 0.0734 | 332.8 | −20.74 | 0.53 | 153±26 | 6974 | 41.99 | 8.03 | −1.77 |

TABLE 1: The Sample—*Continued*

| Source Name | Alternate Name | $C$ | Type | $q$ | $i$ (°) | $D_{25}$ (″) | $z$ | $D_L$ (Mpc) | $M_g$ (mag) | $g-r$ (mag) | $\sigma_* \pm$ error (km s$^{-1}$) | FWHM$_{\mathrm{H}\alpha}$ (km s$^{-1}$) | $\log L_{\mathrm{H}\alpha}$ ($L_\odot$) | $\log M_{\mathrm{BH}}$ ($M_\odot$) | $\log L_{\mathrm{bol}}$ ($L_{\mathrm{Edd}}$) |
|---|---|---|---|---|---|---|---|---|---|---|---|---|---|---|---|
| (1) | (2) | (3) | (4) | (5) | (6) | (7) | (8) | (9) | (10) | (11) | (12) | (13) | (14) | (15) | (16) |
| SDSS J113111.94+100231.3 | ⋯ | 0.46 | Sa | 0.48 | 66.9 | 12.7 | 0.0744 | 337.3 | −21.05 | 0.37 | 104±13 | 2024 | 41.01 | 6.39 | −0.98 |
| SDSS J113249.28+101747.3 | IC 2921 | 0.34 | SBa | 0.43 | 70.7 | 17.1 | 0.0437 | 193.6 | −20.81 | 0.55 | 210±32 | 4002 | 41.64 | 7.34 | −1.38 |
| SDSS J114008.71+030711.3 | ⋯ | 0.41 | SBa | 0.98 | 11.9 | 7.5 | 0.0811 | 369.5 | −20.67 | 0.33 | 92±39 | 591 | 41.33 | 5.46 | 0.23 |
| SDSS J114105.71+024117.0 | | 0.23 | E | 0.74 | 44.9 | 10.7 | 0.0931 | 427.9 | −21.75 | 0.27 | (283±28) | 2423 | 42.61 | 7.43 | −0.63 |
| SDSS J115038.86+020854.2 | | 0.36 | Sa | 0.33 | 82.2 | 14.1 | 0.1094 | 508.2 | −20.84 | 0.77 | 155±36 | 6013 | 41.32 | 7.53 | −1.85 |
| SDSS J120011.27+100135.6 | VIII Zw 160 | 0.32 | E | 0.92 | 24.1: | 8.2 | 0.0854 | 390.1 | −21.73 | 0.25 | 123±28 | 3787 | 42.38 | 7.70 | −1.10 |
| SDSS J120257.81+045045.0 | IC 756 | 0.49 | SBcd | 0.40 | 74.1 | 58.2 | 0.0207 | 90.1 | −20.75 | 0.50 | 102±20 | 2252 | 40.12 | 5.99 | −1.35 |
| SDSS J120330.23+123140.9 | | 0.40 | Sb | 0.63 | 54.3 | 13.4 | 0.0646 | 290.8 | −20.97 | 0.42 | 75±15 | 2511 | 40.56 | 6.33 | −1.31 |
| SDSS J120332.94+022934.5 | UM 472 | 0.24 | SBc | 0.70 | 48.8 | 18.8 | 0.0775 | 352.1 | −22.29 | 0.15 | 125±31 | 2666 | 42.26 | 7.32 | −0.83 |
| SDSS J120648.31+065912.2 | | 0.30 | E | ⋯ | ⋯ | ⋯ | 0.0803 | 365.5 | −20.71 | 0.56 | 133±31 | 4135 | 41.57 | 7.33 | −1.43 |
| SDSS J121629.91+084253.3 | | 0.32 | S0 | 0.41 | 72.6 | 9.4 | 0.0709 | 320.5 | −20.42 | 0.64 | 118±29 | 3391 | 41.85 | 7.31 | −1.17 |
| SDSS J121930.87+064334.4 | MS 1217.0+0700 | 0.35 | Sa | 0.96 | 16.7: | 7.6 | 0.0804 | 366.1 | −21.78 | 0.25 | 113±23 | 1671 | 42.30 | 6.93 | −0.40 |
| SDSS J122042.00+112405.2 | | 0.41 | Sb | 0.73 | 45.8 | 9.4 | 0.0932 | 428.1 | −21.55 | 0.48 | 131±45 | 3155 | 41.76 | 7.20 | −1.14 |
| SDSS J122324.13+024044.4 | Mrk 50 | 0.34 | E/S0 | 0.59 | 57.8: | 22.5 | 0.0236 | 102.9 | −19.84 | 0.55 | 78±15 | 3878 | 41.64 | 7.32 | −1.36 |
| SDSS J122811.42+095126.8 | ⋯ | 0.34 | E/S0 | 0.87 | 31.7: | 11.5 | 0.0639 | 287.4 | −20.78 | 0.61 | 123±21 | 4888 | 41.09 | 7.22 | −1.74 |
| SDSS J123235.82+060310.0 | ⋯ | 0.26 | E | 0.97 | 14.8: | 7.6 | 0.0838 | 382.5 | −21.24 | 0.27 | 175±30 | 2679 | 41.98 | 7.17 | −0.92 |
| SDSS J123639.78+040758.4 | ⋯ | 0.33 | S0 | 0.28 | 90.0 | 20.4 | 0.0706 | 319.0 | −21.06 | 0.54 | 147±22 | 5808 | 41.30 | 7.49 | −1.83 |
| SDSS J123944.23+120159.9 | | 0.43 | Sb | 0.55 | 61.3 | 13.4 | 0.0857 | 391.7 | −21.72 | 0.71 | 223±30 | 6360 | 40.90 | 7.35 | −2.03 |
| SDSS J124239.13+054717.6 | ⋯ | 0.36 | SBa | 0.84 | 35.0: | 7.5 | 0.0936 | 430.1 | −20.73 | 0.38 | 107±29 | 2895 | 41.27 | 6.85 | −1.21 |
| SDSS J124319.98+025256.2 | | 0.35 | S0/Sp | 0.78 | 40.8 | 6.9 | 0.0867 | 396.6 | −20.89 | 0.57 | 122±15 | 1084 | 41.69 | 6.20 | −0.20 |
| SDSS J124635.24+022208.7 | PG 1244+026 | 0.32 | E | 0.95 | 19.2: | 7.2 | 0.0482 | 214.3 | −20.85 | −0.04 | (171±17) | 966 | 41.73 | 6.12 | −0.09 |
| SDSS J124913.75+151510.5 | | 0.35 | Sa | 0.80 | 39.3 | 12.6 | 0.0834 | 380.6 | −21.64 | 0.80 | 197±22 | 2437 | 41.31 | 6.72 | −1.05 |
| SDSS J125039.08+141241.0 | | 0.33 | Sa | 0.46 | 68.2 | 15.9 | 0.0847 | 386.7 | −21.43 | 0.77 | 177±20 | 2682 | 41.44 | 6.88 | −1.10 |
| SDSS J130241.53+040738.6 | ⋯ | 0.40 | S0/Sp | 0.97 | 14.6: | 3.7 | 0.1022 | 472.3 | −20.50 | 0.21 | 186±40 | 4819 | 41.81 | 7.60 | −1.50 |
| SDSS J130803.04+035114.5 | ⋯ | 0.30 | E | 0.69 | 49.3 | 9.6 | 0.0706 | 319.4 | −20.87 | 0.28 | 161±52 | 981 | 41.80 | 6.17 | −0.07 |
| SDSS J132026.49+051113.5 | ⋯ | 0.35 | S0 | 0.64 | 53.9 | 6.7 | 0.0984 | 453.7 | −20.96 | 0.14 | 102±28 | 1617 | 41.89 | 6.67 | −0.50 |
| SDSS J132442.44+052438.8 | ⋯ | 0.41 | S0/Sp | 0.79 | 40.0 | 3.6 | 0.1158 | 540.6 | −20.08 | 0.06 | 127±70 | 12025 | 41.93 | 8.49 | −2.28 |
| SDSS J134952.84+020445.1 | UM 614 | 0.27 | S0 | 0.51 | 64.5 | 21.0 | 0.0328 | 144.4 | −20.35 | 0.39 | 138±13 | 2657 | 41.71 | 7.02 | −1.00 |
| SDSS J140018.42+050242.2 | TOL 1358+052 | 0.41 | SBb | 0.76 | 43.3 | 19.0 | 0.0342 | 150.5 | −21.29 | 0.30 | 145±14 | 3340 | 41.26 | 6.97 | −1.34 |
| SDSS J142748.28+050222.0 | ⋯ | 0.46 | E/S0 | 0.96 | 17.2: | 4.8 | 0.1061 | 491.8 | −22.45 | −0.27 | (26±3) | 1513 | 42.62 | 7.01 | −0.21 |
| SDSS J143450.62+033842.5 | ⋯ | 0.53 | Scd | 0.86 | 32.2 | 16.3 | 0.0286 | 123.6 | −20.19 | 0.27 | 77±25 | 1480 | 41.87 | 6.58 | −0.42 |
| SDSS J145047.19+033645.4 | ⋯ | 0.40 | S0/Sp | 0.56 | 60.7: | 5.0 | 0.0665 | 300.0 | −19.34 | 0.31 | 119±21 | 1987 | 41.05 | 6.39 | −0.94 |
| SDSS J150556.55+034226.3 | Mrk 1392 | 0.27 | SBb | 0.53 | 62.8 | 33.4 | 0.0359 | 158.1 | −22.07 | 0.16 | 199±18 | 4373 | 42.12 | 7.69 | −1.32 |
| SDSS J153839.19+024324.7 | | 0.39 | Sb | 0.69 | 49.2 | 20.3 | 0.0314 | 137.9 | −18.98 | 0.28 | (953±95) | 3982 | 40.50 | 6.71 | −1.74 |
| SDSS J155417.43+323837.8 | | 0.34 | S0/Sp | 0.83 | 36.0 | 9.8 | 0.0483 | 214.9 | −21.04 | 0.55 | 122±32 | 4734 | 42.10 | 7.75 | −1.40 |
| SDSS J160936.42+254459.6 | | 0.35 | S0/Sp | 0.89 | 29.1: | 7.6 | 0.0411 | 181.8 | −19.97 | 0.66 | 145±22 | 8562 | 41.40 | 7.89 | −2.14 |
| SDSS J163152.23+345328.7 | | 0.41 | SBa | 0.85 | 33.8 | 11.7 | 0.0732 | 331.6 | −21.54 | 0.56 | 130±21 | 1633 | 41.73 | 6.59 | −0.56 |
| SDSS J163159.59+243740.3 | | 0.37 | E/S0 | 0.89 | 29.0: | 5.7 | 0.0435 | 192.7 | −19.39 | 0.47 | 82±26 | 836 | 40.84 | 5.50 | −0.23 |
| SDSS J163453.67+231242.7 | MCG +04-39-016 | 0.45 | Sb | 0.78 | 40.7 | 16.5 | 0.0386 | 170.5 | −21.50 | 0.51 | 117±20 | 4354 | 40.96 | 7.05 | −1.68 |
| SDSS J163523.44+263046.5 | | 0.31 | E/S0 | 0.64 | 53.9 | 10.3 | 0.0708 | 319.8 | −20.89 | 0.76 | 107±18 | 2501 | 41.14 | 6.65 | −1.13 |
| SDSS J165601.61+211241.2 | | 0.36 | Sp | 0.50 | 65.2: | 9.7 | 0.0491 | 218.6 | −20.18 | 0.68 | 109±29 | 5429 | 41.50 | 7.54 | −1.70 |
| SDSS J170102.29+340400.5 | | 0.38 | E/S0 | 0.69 | 49.8: | 4.5 | 0.0944 | 433.9 | −20.14 | 0.53 | 168±37 | 5917 | 42.47 | 8.15 | −1.48 |
| SDSS J171257.02+231526.8 | | 0.38 | Sa | 0.47 | 67.6 | 11.2 | 0.0570 | 255.2 | −20.62 | 0.73 | 141±21 | 5243 | 40.61 | 7.02 | −1.95 |
| SDSS J171520.19+274825.9 | | 0.36 | Sa | 0.45 | 69.5 | 6.7 | 0.1069 | 496.2 | −20.55 | 0.61 | 59±27[b] | 5023 | 41.67 | 7.56 | −1.58 |
| SDSS J171601.93+311213.8 | | 0.41 | Sa | 0.94 | 21.0 | 8.9 | 0.1102 | 512.1 | −23.07 | 0.09 | (367±37) | 1648 | 43.20 | 7.41 | −0.10 |
| SDSS J210622.99+105606.5 | ⋯ | 0.39 | Sb | 0.80 | 39.4 | 6.5 | 0.0947 | 435.8 | −21.18 | −0.12 | 106±28 | 3179 | 41.36 | 6.98 | −1.26 |

TABLE 1: THE SAMPLE—*Continued*

| Source Name | Alternate Name | $C$ | Type | $q$ | $i$ (°) | $D_{25}$ ('') | $z$ | $D_L$ (Mpc) | $M_g$ (mag) | $g-r$ (mag) | $\sigma_* \pm$ error (km s$^{-1}$) | FWHM$_{H\alpha}$ (km s$^{-1}$) | log $L_{H\alpha}$ ($L_\odot$) | log $M_{BH}$ ($M_\odot$) | log $L_{bol}$ ($L_{Edd}$) |
|---|---|---|---|---|---|---|---|---|---|---|---|---|---|---|---|
| (1) | (2) | (3) | (4) | (5) | (6) | (7) | (8) | (9) | (10) | (11) | (12) | (13) | (14) | (15) | (16) |
| SDSS J210721.91+110359.0 | ⋯ | 0.37 | S0/Sp | 0.42 | 71.7 | 9.2 | 0.0423 | 187.1 | −19.83 | −0.27 | 69±23 | 4108 | 40.81 | 6.91 | −1.67 |
| SDSS J211646.34+110237.4 | ⋯ | 0.34 | Sa | 0.92 | 24.5 | 12.3 | 0.0806 | 366.9 | −22.01 | −0.18 | 104±48 | 2861 | 42.21 | 7.36 | −0.91 |
| SDSS J223338.42+131243.6 | ⋯ | 0.34 | SB0 | 0.52 | 63.4 | 14.2 | 0.0934 | 429.4 | −22.26 | 0.06 | 161±46 | 3122 | 42.24 | 7.45 | −0.97 |
| SDSS J225305.11+140351.2 | ⋯ | 0.32 | E/S0 | 0.74 | 44.7: | 8.1 | 0.0925 | 424.9 | −20.93 | 0.26 | (115±12) | 4250 | 41.44 | 7.29 | −1.51 |
| SDSS J232721.96+152437.3 | ⋯ | 0.28 | E | 0.73 | 46.0 | 23.0 | 0.0457 | 203.0 | −21.84 | 0.37 | (133±13) | 7541 | 41.69 | 7.94 | −1.94 |
| SDSS J235128.78+155259.0 | ⋯ | 0.35 | S0/Sp | 0.58 | 58.8 | 7.2 | 0.0963 | 443.4 | −21.52 | 0.63 | 185±16 | 7881 | 42.24 | 8.28 | −1.80 |
| 3C 120 | Mrk 1506 | ⋯ | S0 | 0.62 | 55.3 | 46.6 | 0.0330 | 145.7 | −22.50 | ⋯ | 162±24 | 2205 | 42.92 | 7.48 | −0.42 |
| Akn 120 | Mrk 1095 | ⋯ | E/S0 | 0.81 | 37.9 | 56.0 | 0.0327 | 144.3 | −22.25 | ⋯ | 221±17 | 5410 | 42.99 | 7.92 | −0.80 |
| MCG +01−13−012 | UGC 3223 | ⋯ | SBa | 0.44 | 70.3 | 92.9 | 0.0156 | 67.1 | −20.40 | ⋯ | 106±28 | 4740 | 40.74 | 7.20 | −2.02 |
| NGC 3227 | ⋯ | ⋯ | SABa | 0.51 | 64.4 | 244.4 | 0.00386 | 17.1 | −20.16 | ⋯ | 136±4 | 5208 | 40.85 | 7.37 | −2.10 |
| NGC 5548 | Mrk 1509 | ⋯ | S0/a | 0.79 | 40.0 | 80.9 | 0.0172 | 74.3 | −21.14 | ⋯ | 201±12 | 5662 | 42.15 | 7.57 | −1.17 |
| NGC 7469 | Mrk 1514 | ⋯ | SABa | 0.83 | 35.8 | 82.8 | 0.0163 | 71.4 | −21.57 | ⋯ | 131±5 | 2169 | 42.40 | 6.83 | −0.22 |
| PG 0003+199 | Mrk 335 | ⋯ | E | ⋯ | ⋯ | ⋯ | 0.0258 | 112.9 | −21.27 | ⋯ | ⋯ | 1502 | 42.56 | 6.89 | −0.14 |
| PG 0844+349 | TON 951 | ⋯ | Sp(d) | ⋯ | ⋯ | ⋯ | 0.0640 | 287.6 | −22.18 | ⋯ | ⋯ | 2148 | 43.13 | 7.71 | −0.47 |
| PG 1211+143 | ⋯ | ⋯ | ? | ⋯ | ⋯ | ⋯ | 0.0809 | 368.6 | −23.74 | ⋯ | ⋯ | 1317 | 43.59 | 7.90 | −0.26 |
| PG 1229+204 | Mrk 771 | ⋯ | SB0 | 0.64 | 53.7 | 6.0 | 0.0630 | 282.9 | −22.09 | ⋯ | ⋯ | 3415 | 42.81 | 7.60 | −0.63 |
| PG 1307+085 | ⋯ | ⋯ | ? | ⋯ | ⋯ | ⋯ | 0.1550 | 742.9 | −23.72 | ⋯ | ⋯ | 5058 | 43.74 | 8.38 | −0.61 |
| PG 1426+015 | Mrk 1383 | ⋯ | E/S0 | ⋯ | ⋯ | 4.4 | 0.0865 | 395.7 | −23.31 | ⋯ | ⋯ | 6323 | 43.55 | 8.85 | −1.24 |
| PG 2130+099 | II Zw 136 | ⋯ | S0 | 0.64 | 53.7 | 10.4 | 0.0629 | 282.9 | −22.58 | ⋯ | ⋯ | 2912 | 43.25 | 8.40 | −1.05 |
| RX J0602.1+2828 | ⋯ | ⋯ | ? | ⋯ | ⋯ | ⋯ | 0.033 | 145.7 | −22.7 | ⋯ | ⋯ | 7117 | 41.80 | 8.15 | −2.05 |
| RX J0608.0+3058 | ⋯ | ⋯ | ? | ⋯ | ⋯ | ⋯ | 0.073 | 330.0 | −23.6 | ⋯ | ⋯ | 5918 | 44.29 | 9.38 | −1.13 |

NOTE.— Col. (1) Source name. Col. (2) Alternate name. Col. (3) The (inverse) concentration index in the $g$ band, defined as the ratio of Petrosian radii enclosing 50% to 90% of the light. Col. (4) Adopted morphological type. For SDSS objects, this was estimated either from $C$, following Shimasaku et al. 2001, or from visual inspection of the images presented in Fig. 1; for non-SDSS objects, this was either taken from the RC3, or estimated from visual inspection of *HST* images, where available. See text. We use conventional notation for most objects, except in cases where we cannot distinguish a true elliptical vs. an S0 (E/S0) or between a true S0 vs. a spiral (S0/Sp). "Sp(d)" denotes a tidally disturbed spiral. Col. (5) Ratio of minor axis to major axis ($q = b/a$), measured at a surface brightness level of $\mu = 25$ mag arcsec$^{-2}$ in the $g$ and $B$ band for SDSS and non-SDSS (from Hyperleda) objects, respectively. Col. (6) Inclination angle, calculated from $q$ using Hubble's 1926 formula assuming $q_0 = 0.3$ (Fouqué et al. 1990). Values that are deemed suspect or particularly uncertain are followed by a colon. The values for SDSS J080538.66+261005.4 and SDSS J080546.97+260532.9 were estimated visually. Col. (7) Isophotal diameter, measured at a surface brightness level of $\mu = 25$ mag arcsec$^{-2}$ in the $g$ and $B$ band for SDSS and non-SDSS (from Hyperleda) objects, respectively. Col. (8) Optical redshift from the NASA/IPAC Extragalactic Database. Col. (9) Luminosity distance calculated assuming $H_0 = 70$ km s$^{-1}$ Mpc$^{-1}$, $\Omega_m = 0.27$, and $\Omega_\Lambda = 0.73$. Col. (10) Total (host galaxy + AGN) absolute magnitude, corrected for Galactic extinction; SDSS objects are given in the $g$ band and non-SDSS objects in the $B$ band. Col. (11) Total (host galaxy + AGN) $g - r$ color, corrected for Galactic extinction. Col. (12) Central velocity dispersion, derived either from stars (Barth et al. 2005; Greene & Ho 2006b; this work) or from the width of the [O II] $\lambda 3727$ line (values given in parentheses), as described in Greene & Ho 2005a. Additional data come from Onken et al. 2004 (3C 120), Nelson et al. 2004 (Akn 120, NGC 3227, NGC 5548, NGC 7469), and Nelson & Whittle 1995 (MCG +01−13−012). Col. (13) FWHM of the broad H$\alpha$ line for the SDSS objects and of the broad H$\beta$ line for the rest. Data for the non-SDSS objects come from Peterson et al. 2004, except for MCG +01−13−012 (Stirpe 1990) and RX J0602.1+2828 and RX J0608.0+3058 (Motch et al. 1998, measured from the published spectra, which were kindly sent to us by C. Motch). Col. (14) Luminosity of the broad H$\alpha$ line. For the non-SDSS objects, the H$\alpha$ luminosity was estimated from the 5100 Å continuum luminosity of Peterson et al. 2004, using the $L_{H\alpha} - L_{5100\text{Å}}$ correlation (Eq. 1) of Greene & Ho 2005b, with the exception of MCG +01−13−012 (Stirpe 1990) and RX J0602.1+2828 and RX J0608.0+3058 (Motch et al. 1998), whose published H$\beta$ luminosities were converted to H$\alpha$ assuming H$\alpha$/H$\beta$ = 3.5 (see Greene & Ho 2005b). Col. (15) Black hole mass derived from the H$\alpha$ method (Eq. 6) of Greene & Ho 2005b, except for the reverberation-mapped objects, whose masses were taken directly from Peterson et al. 2004, and MCG +01−13−012, RX J0602.1+2828, RX J0608.0+3058, whose masses are based on the H$\beta$ method (Eq. 7) of Greene & Ho 2005b. For consistency with the virial mass zeropoint adopted by Greene & Ho 2005b, all the reverberation-mapped masses were reduced by a factor of 1.8 (see footnote 4 in Greene & Ho 2005b). Col. (16) Eddington ratio, assuming $L_{bol} = 9.8 L_{5100\text{Å}}$ (McLure & Dunlop 2004), with $L_{5100\text{Å}}$ estimated from the $L_{H\alpha} - L_{5100\text{Å}}$ correlation of Greene & Ho 2005b.

[a] Taken from Greene & Ho (2006a).
[b] Possibly unreliable.

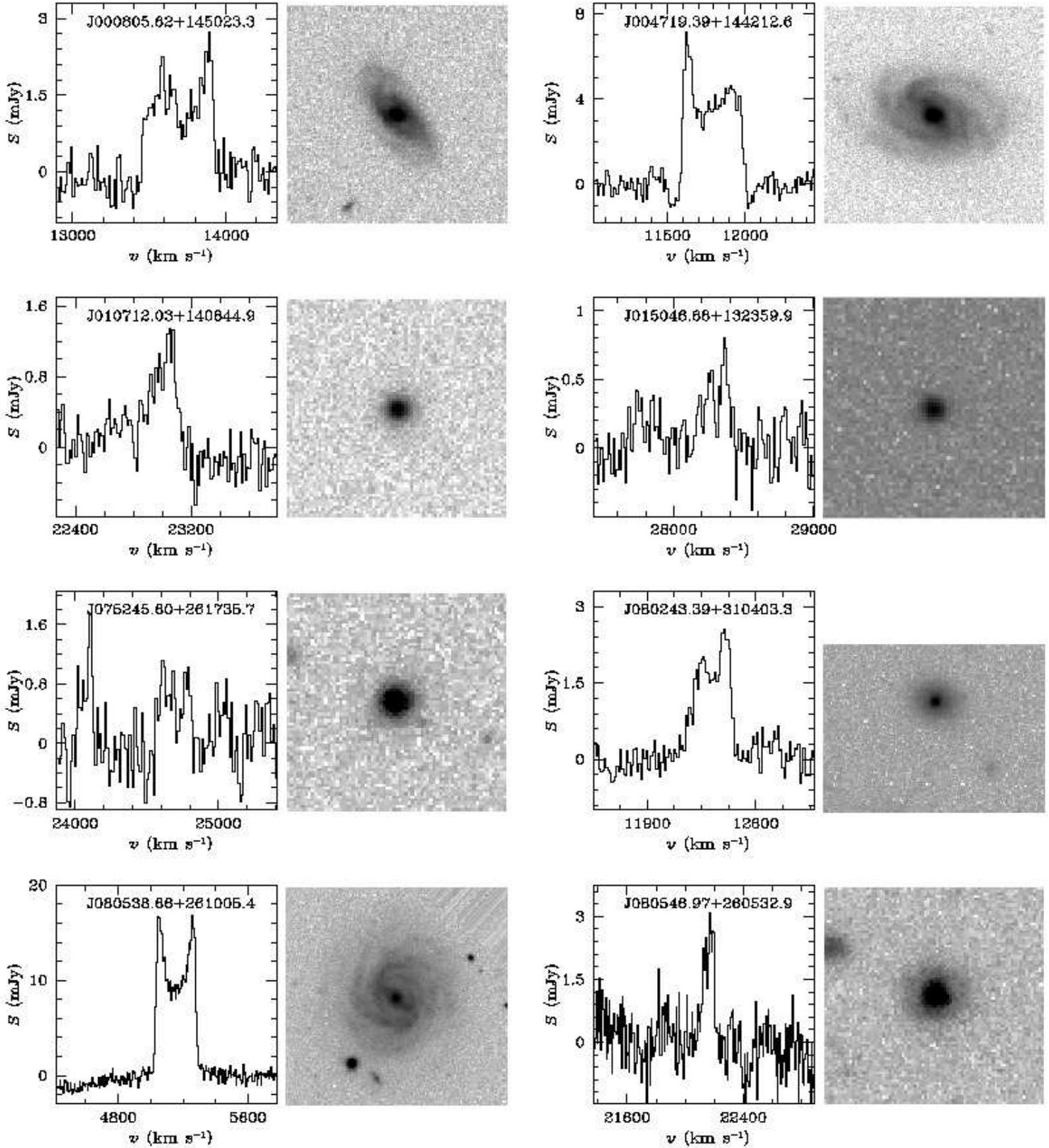

FIG. 3a.— H I spectra and optical *g*-band SDSS images of the H I-detected objects. The velocity scale is given in the barycentric frame, and the velocity range is chosen such that the lines have roughly comparable widths on the plots. Features suspected to be due to radio frequency interference are labeled "RFI." Each image subtends a physical scale of 50 kpc × 50 kpc, with north oriented up and east to the left.

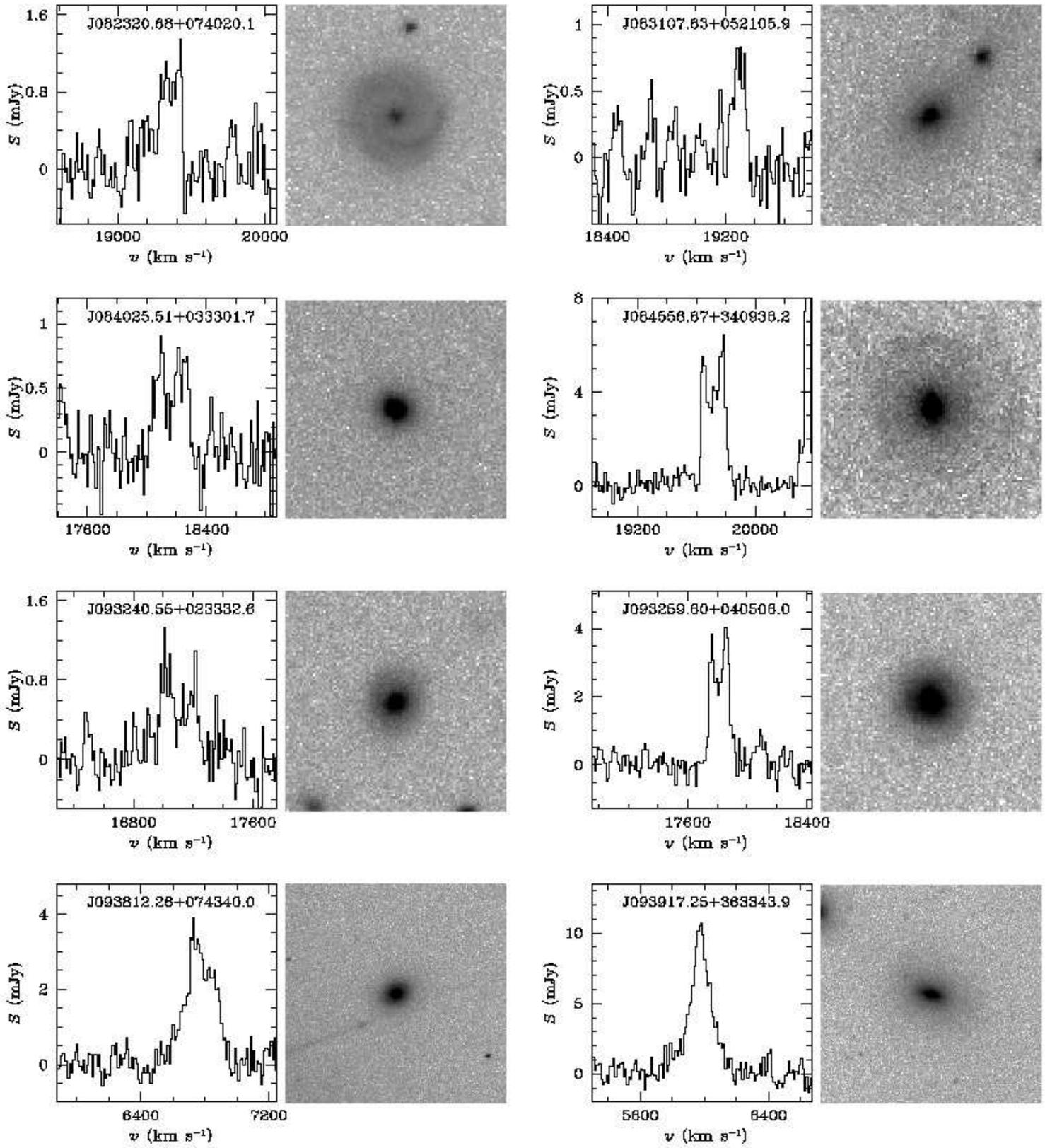

FIG. 3b.— Same as Fig. 3a.

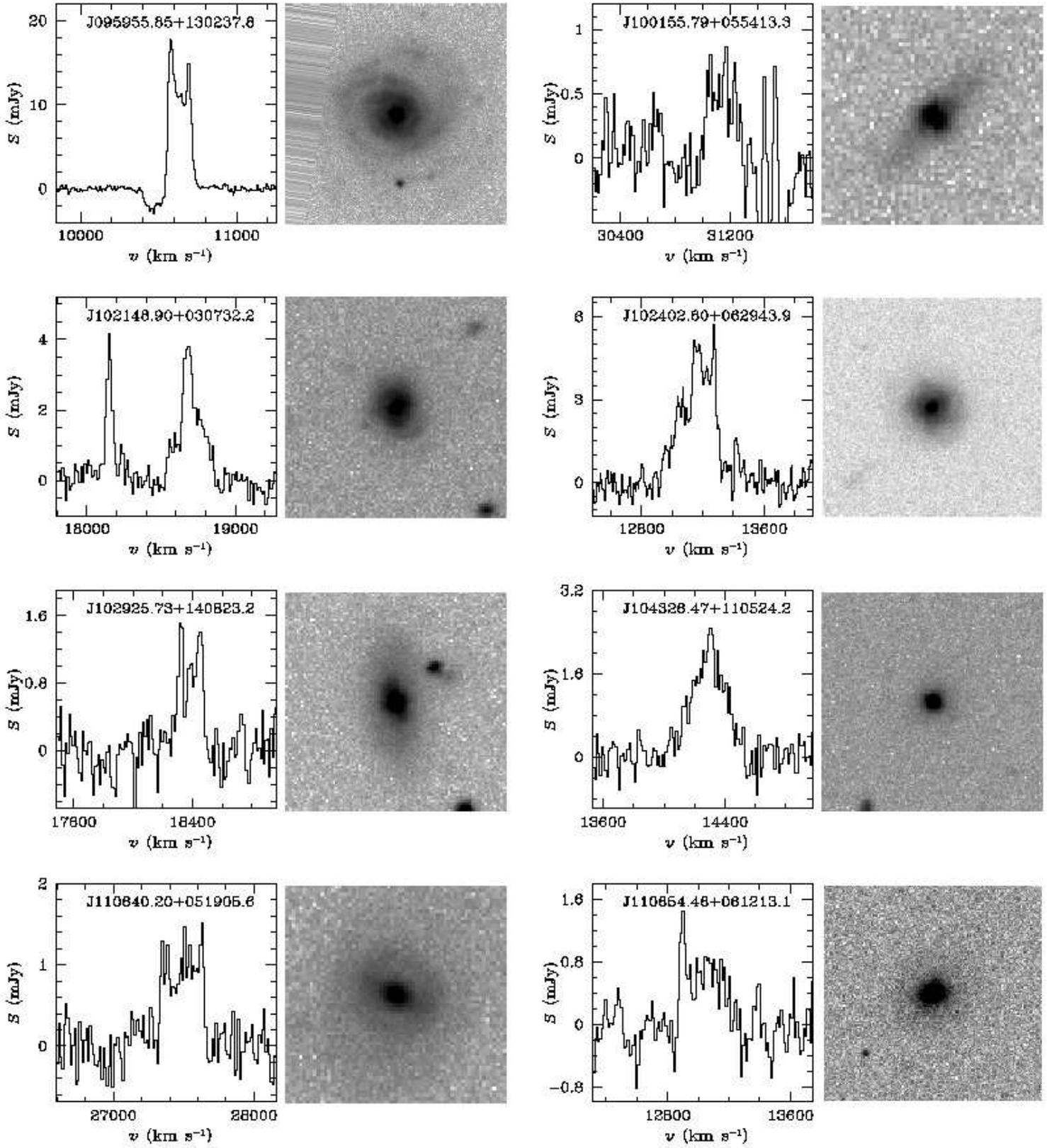

Fig. 3*c*.— Same as Fig. 3*a*.

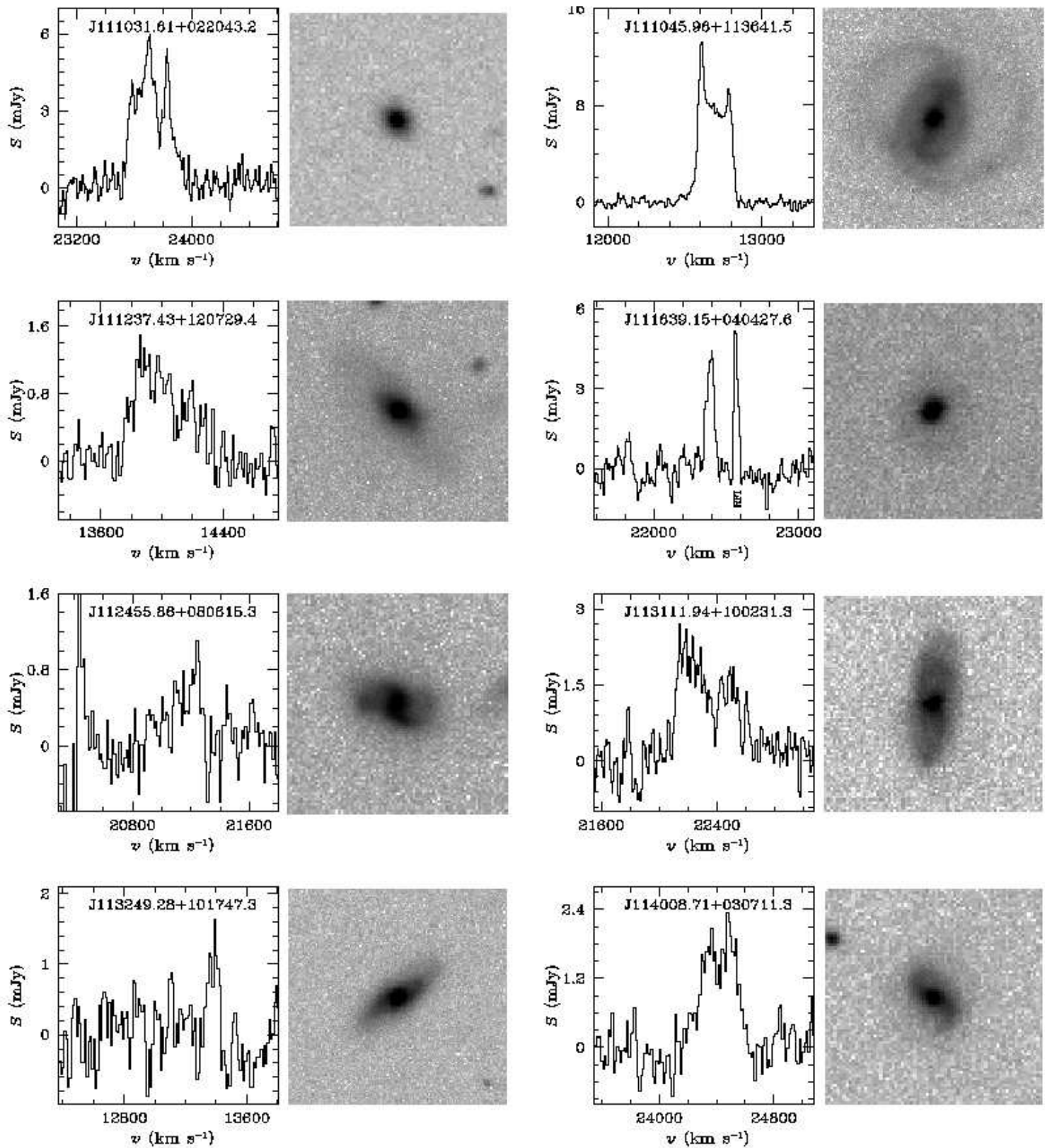

FIG. 3*d*.— Same as Fig. 3*a*.

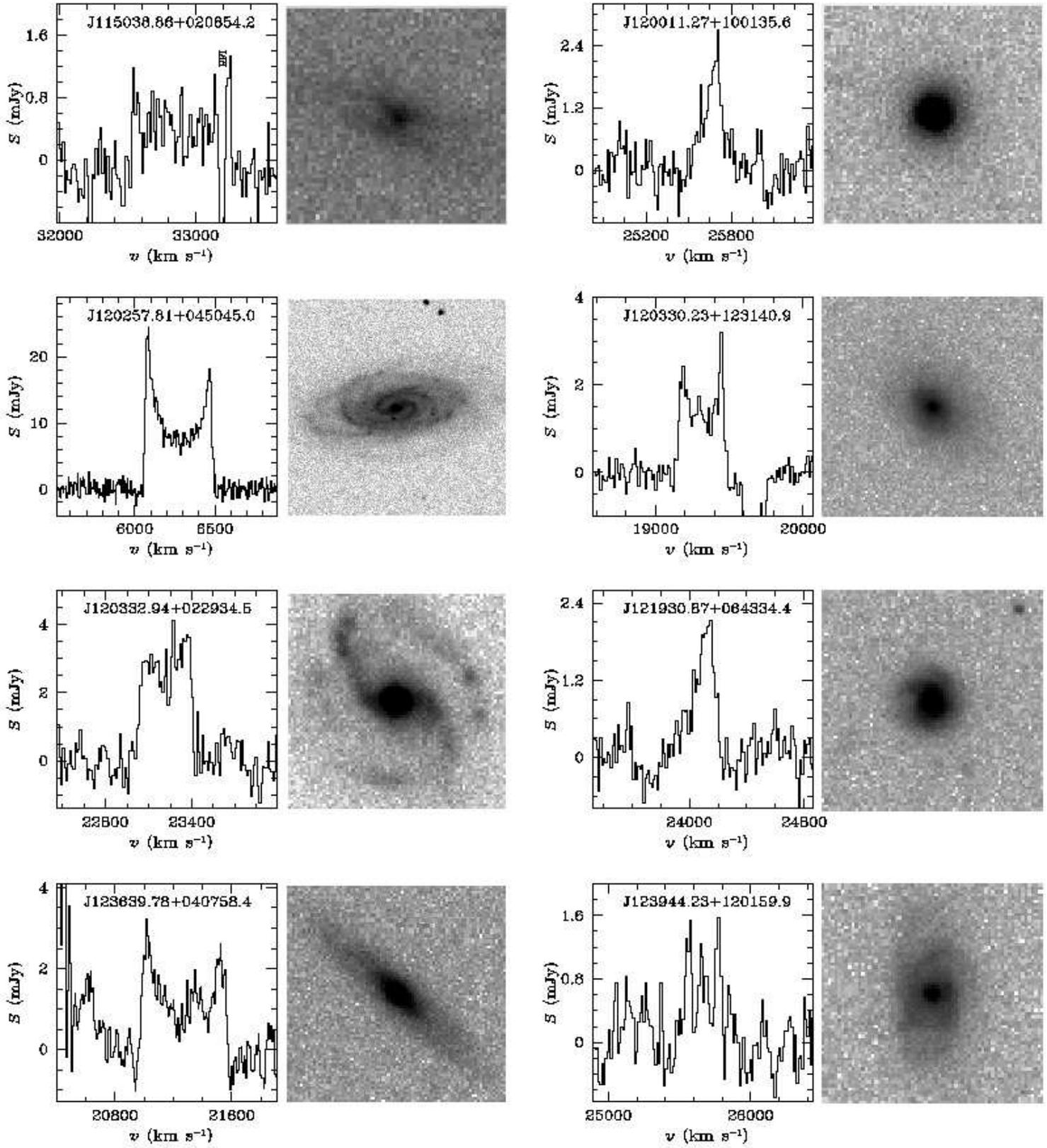

FIG. 3e.— Same as Fig. 3a.

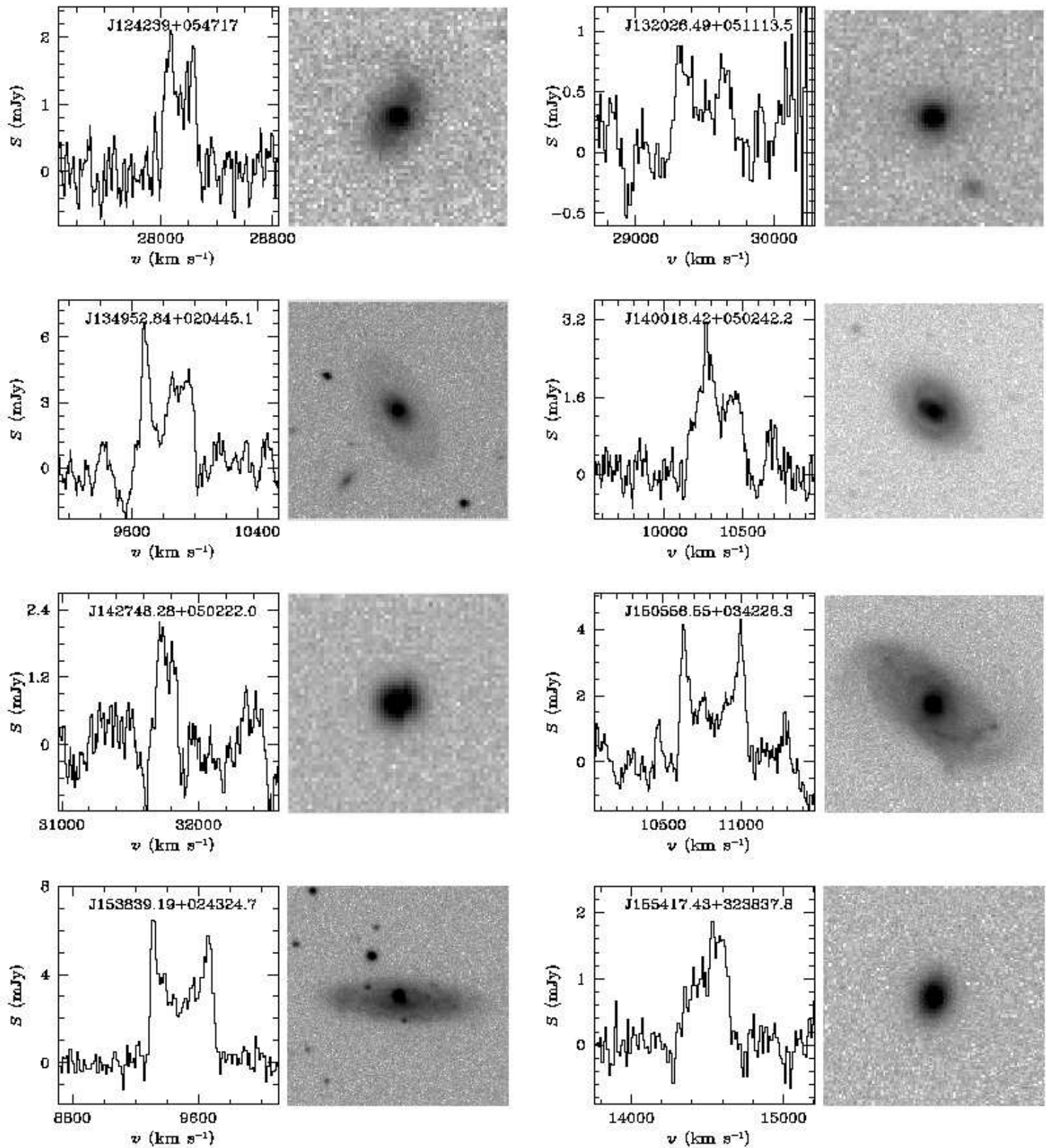

FIG. 3f.— Same as Fig. 3a.

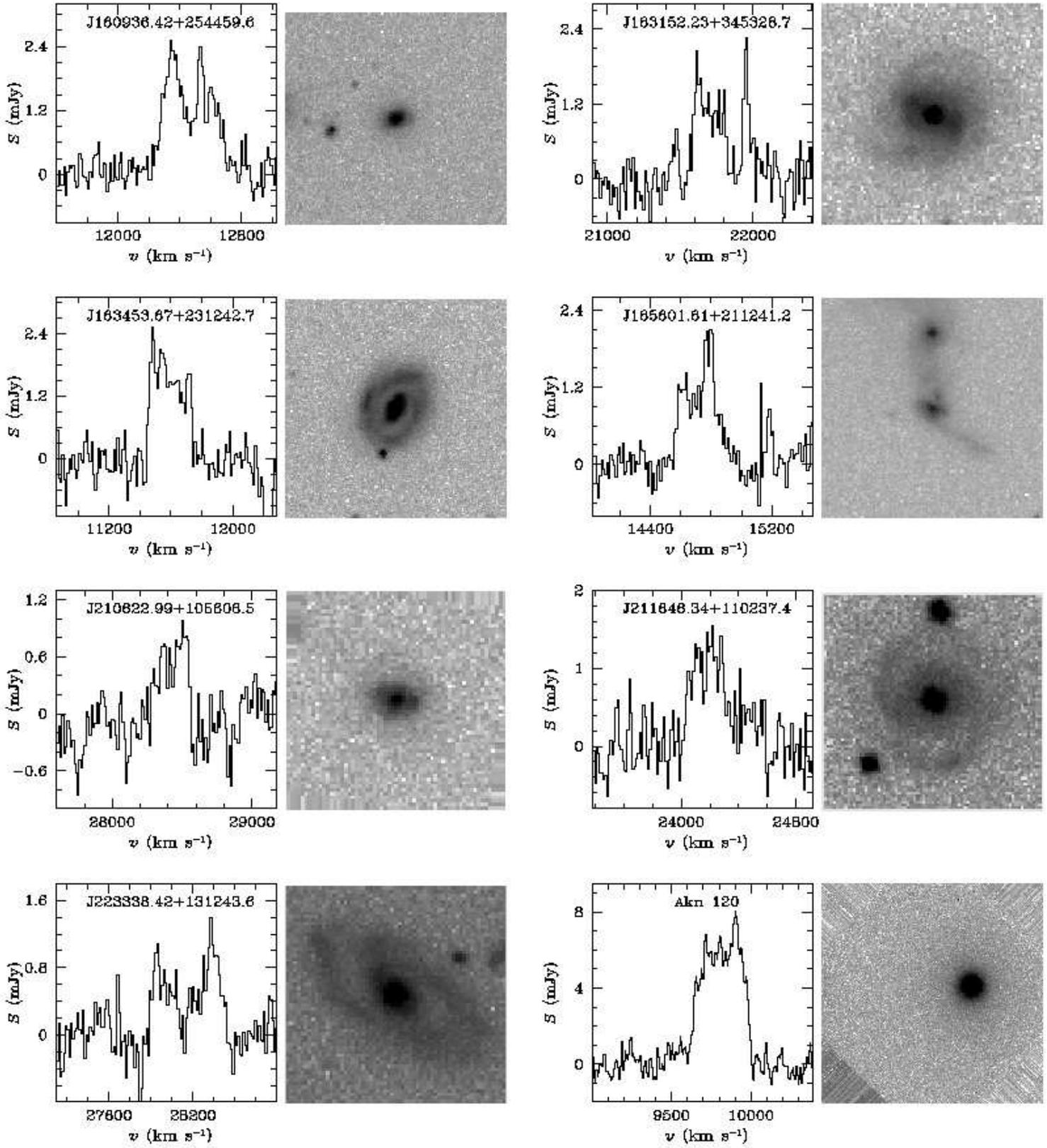

FIG. 3*g*.— Same as Fig. 3*a*. The image for Akn 120 comes from *HST*/PC2 (filter F750LP) and subtends 22.4 kpc × 22.4 kpc.

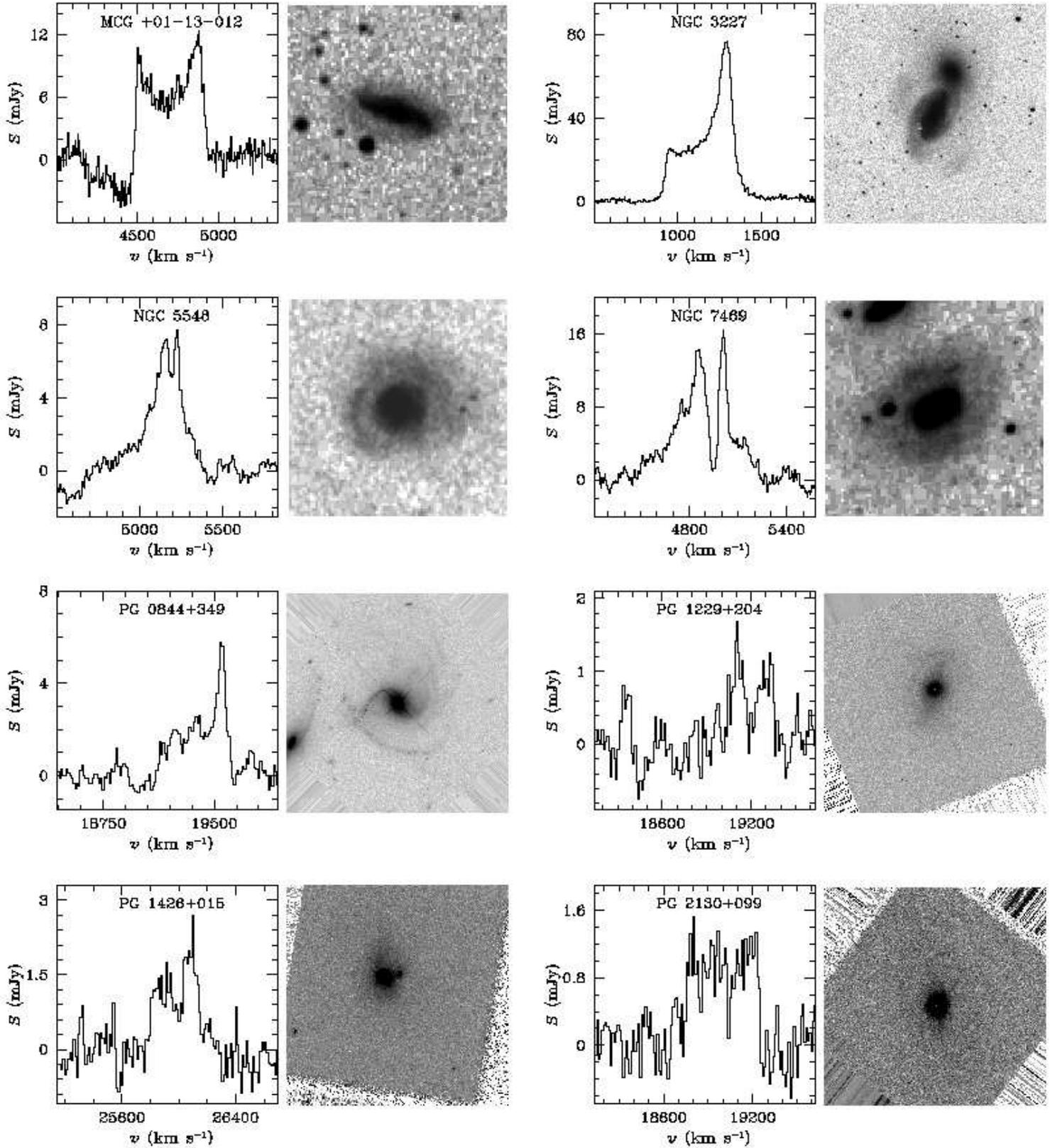

FIG. 3h.— Same as Fig. 3a, except that for MCG +01-13-012, NGC 3227, NGC 5548, and NGC 7469 the images are in the B band taken from the Digital Sky Survey. The images for the four PG quasars come from HST, taken with the following detector and filter combinations and field sizes: PG 0844+349 (ACS/F625W, 84.6 kpc × 84.6 kpc), PG 1229+204 (PC2/F606W, 43.9 kpc × 43.9 kpc), PG 1426+015 (PC2/F814W, 61.4 kpc × 61.4 kpc), and PG 2130+099; (PC2/F450W, 43.9 kpc × 43.9 kpc).

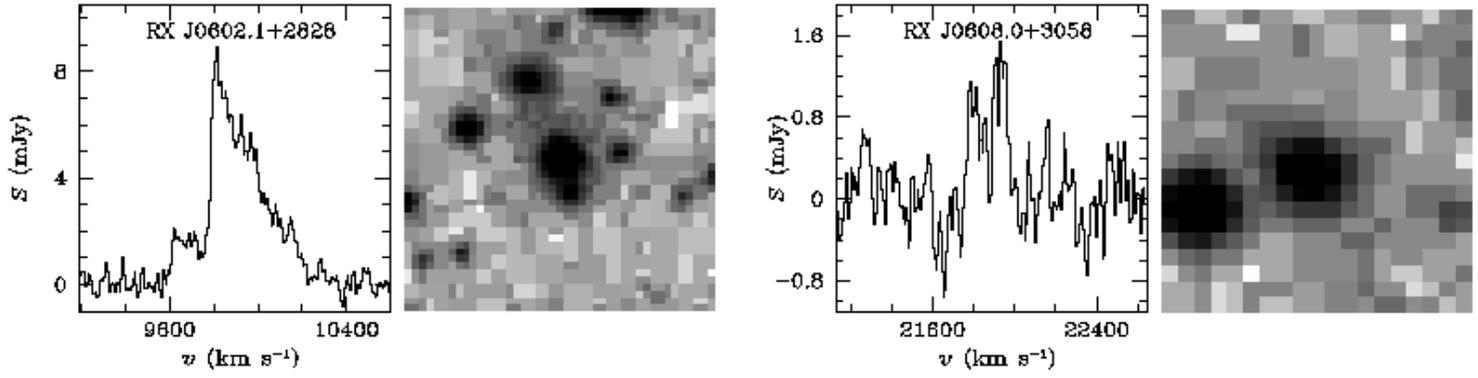

FIG. 3*i*.— Same as Fig. 3*a*, except that for RX J0602.1+2828 and RX J0608.0+3058 the images are in the *B* band and were taken from the Digital Sky Survey.

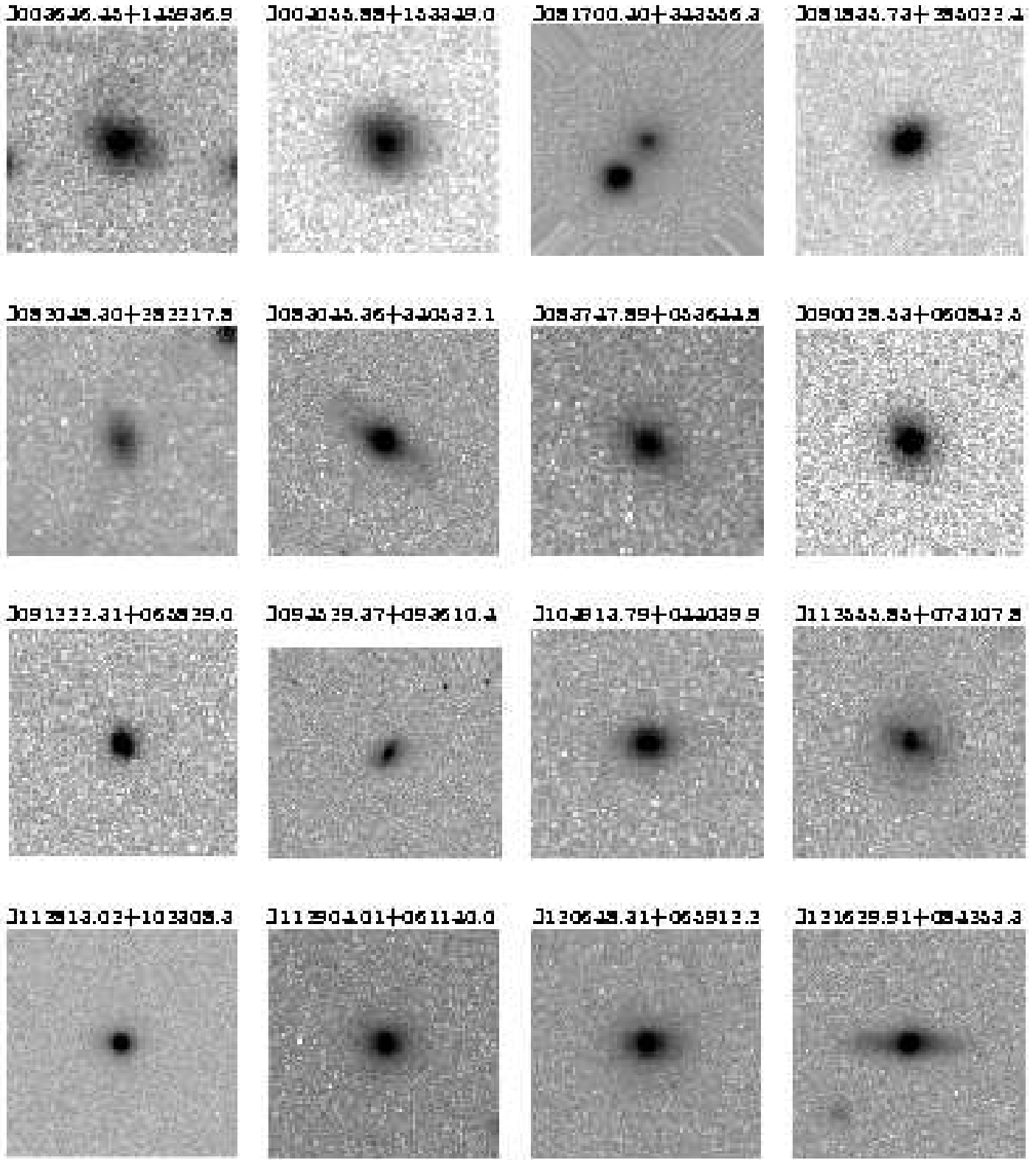

Fig. 4a.— Optical g-band SDSS images of the H I nondetections. Each image subtends a physical scale of 50 kpc × 50 kpc, with north oriented up and east to the left.

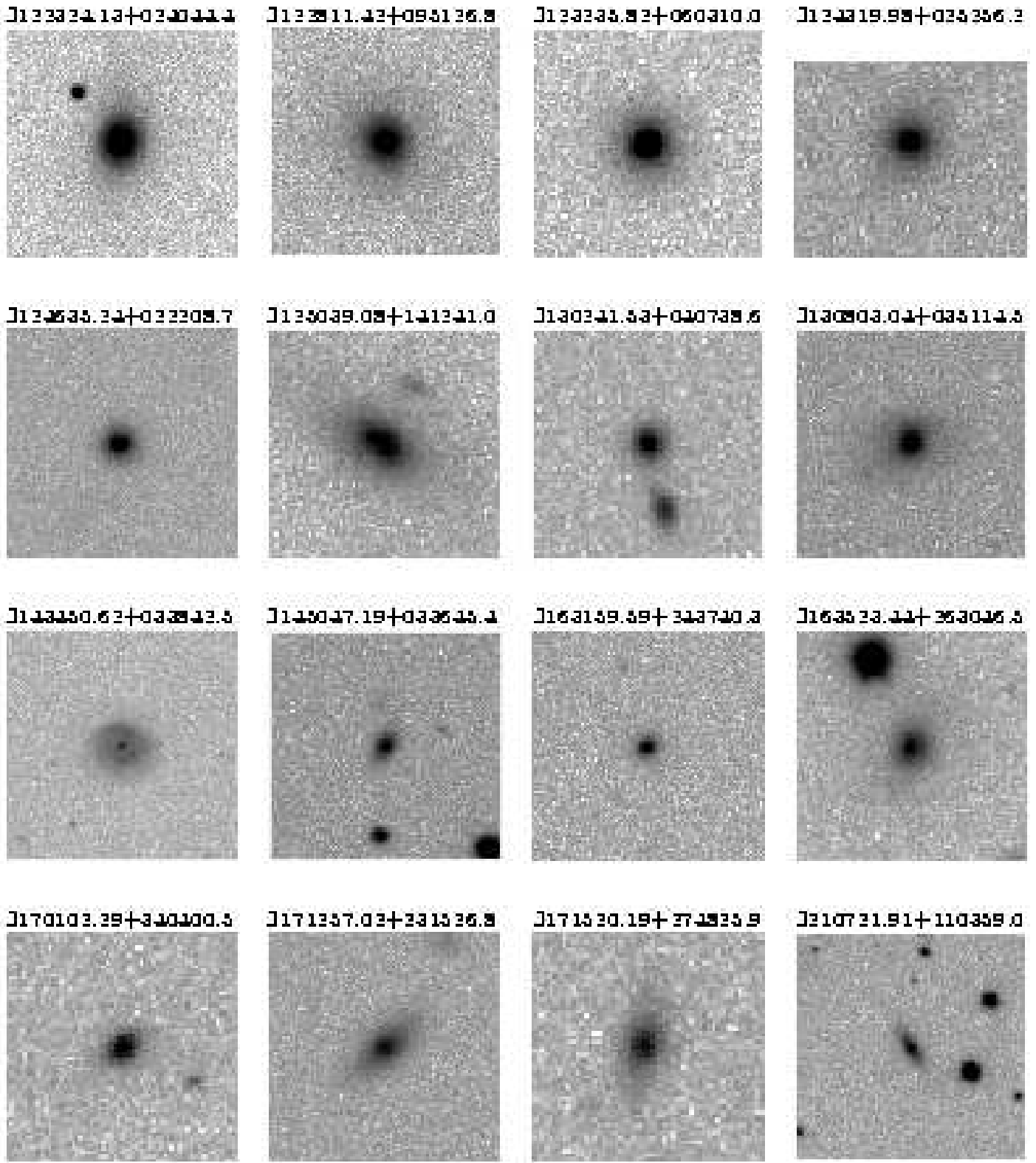

FIG. 4b.— Same as Fig. 4a.

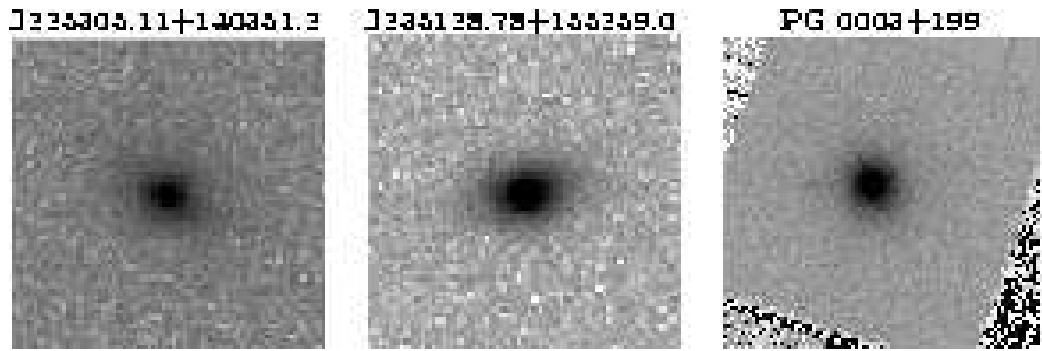

FIG. 4c.— Same as Fig. 4a, except for PG 0003+199, which comes from *HST*/PC2 (filter F606W) and subtends 21.9 kpc × 21.9 kpc.

TABLE 2: H I Properties

| Source Name | Base. | Smooth | $v_{\rm sys}$ (km s$^{-1}$) | $W_{20}$ (km s$^{-1}$) | $v_m$ (km s$^{-1}$) | Prof. | $\int S_\nu dv$ (Jy km s$^{-1}$) | log $M_{\rm H\,I}$ ($M_\odot$) | log $M_{\rm dyn}$ ($M_\odot$) | $M_{B,\rm TF}$ (mag) | Notes |
|---|---|---|---|---|---|---|---|---|---|---|---|
| (1) | (2) | (3) | (4) | (5) | (6) | (7) | (8) | (9) | (10) | (11) | (12) |
| SDSS J000805.62+145023.3 | 5 | Yd | 13688.9 | 477.5 | 227.0 | D | 0.669 | 9.81 | 10.79 | −20.72 | 1,3 |
| SDSS J003646.45+145936.9 | 5 | Yd | ... | ... | ... | ... | <0.33 | <10.11 | ... | ... | 4 |
| SDSS J004055.88+153349.0 | 5 | Yd | ... | ... | ... | ... | <0.21 | <10.01 | ... | ... | 5 |
| SDSS J004719.39+144212.6 | 5 | Yd | 11796.2 | 407.9 | 220.8 | D | 1.608 | 10.06 | 10.86 | −20.63 | 1 |
| SDSS J010712.03+140844.9 | 5 | Yd | 22982.5 | 293.2 | 259.2: | AS | 0.215 | 9.79 | 10.56: | −21.15: | 2,4 |
| SDSS J015046.68+132359.9 | 5 | Yd | 28060.1 | 692.4 | 380.7: | AS | 0.149 | 9.82 | 10.90: | −22.40: | 14,20 |
| SDSS J075245.60+261735.7 | 5 | Yd | 24690.4 | 258.7 | 339.9: | ? | 0.158 | 9.72 | 10.90: | −22.03: | 6 |
| SDSS J080243.39+310403.3 | 5 | Yd | 12300.6 | 313.5 | 259.4 | A | 0.511 | 9.59 | 10.63 | −21.15 | |
| SDSS J080538.66+261005.4 | 5 | N | 5156.2 | 263.8 | 275.1 | D | 2.929 | 9.57 | 11.07 | −21.34 | |
| SDSS J080546.97+260532.9 | 5 | N | 22156.0 | 100.9 | 98.4: | S | 0.196 | 9.71 | 9.86: | −18.00: | |
| SDSS J081700.40+343556.3 | 5 | Yd | ... | ... | ... | ... | <0.47 | <9.94 | ... | ... | 21 |
| SDSS J081835.73+285022.4 | 5 | Yd | ... | ... | ... | ... | <0.35 | <10.00 | ... | ... | |
| SDSS J082048.30+282217.8 | 5 | Yd | ... | ... | ... | ... | <0.26 | <10.06 | ... | ... | |
| SDSS J082320.68+074020.2 | 5 | Yd | 19353.3 | 183.9 | 247.4 | A? | 0.154 | 9.49 | 10.87 | −21.00 | 22 |
| SDSS J083045.36+340532.1 | 5 | Yd | ... | ... | ... | ... | <0.20 | <9.57 | ... | ... | 7,23 |
| SDSS J083107.63+052105.9 | 5 | Yd | 19285.4 | 137.1 | 95.2 | S | 0.074 | 9.15 | 9.82 | −17.89 | 2,24 |
| SDSS J083747.89+053644.8 | 5 | Yd | ... | ... | ... | ... | <0.26 | <10.12 | ... | ... | 7,25 |
| SDSS J084025.51+033301.7 | 5 | Yd | 18159.8 | 263.1 | 198.4: | D | 0.138 | 9.38 | 10.46: | −20.28: | |
| SDSS J084556.67+340936.2 | 5 | Yd | 19717.5 | 192.4 | 159.9 | D | 0.806 | 10.22 | 10.65 | −19.58 | 2 |
| SDSS J090028.53+060842.5 | 5 | Yd | ... | ... | ... | ... | <0.43 | <10.10 | ... | ... | |
| SDSS J091222.31+065829.0 | 5 | Yd | ... | ... | ... | ... | <0.36 | <10.03 | ... | ... | 7 |
| SDSS J093240.55+023332.6 | 5 | Yd | 17103.9 | 290.6 | 188.6 | D | 0.176 | 9.43 | 10.44 | −20.12 | |
| SDSS J093259.60+040506.0 | 5 | Yd | 17818.4 | 162.8 | 130.7 | D | 0.441 | 9.86 | 10.20 | −18.92 | 2 |
| SDSS J093812.26+074340.0 | 5 | Yd | 6776.4 | 304.6 | 174.1 | AS | 0.663 | 9.15 | 10.19 | −19.86 | 2,26 |
| SDSS J093917.26+363343.9 | 5 | Yd | 5978.9 | 219.5 | 119.2 | S | 1.373 | 9.39 | 10.08 | −18.62 | 2 |
| SDSS J094529.37+093610.4 | 5 | Yd | ... | ... | ... | ... | <0.38 | <8.48 | ... | ... | 27 |
| SDSS J095955.85+130237.8 | 5 | Y | 10638.6 | 175.2 | 233.9: | D | 2.052 | 10.07 | 11.13: | −20.81: | 28 |
| SDSS J100155.79+055413.3 | 5 | Yd | 31098.7 | 169.3 | 62.1: | S? | 0.092 | 9.69 | ... | −16.51: | 2,29 |
| SDSS J102148.90+030732.2 | 5 | Yd | 18688.8 | 284.4 | 177.4 | AS | 0.520 | 9.98 | 10.46 | −19.92 | 30 |
| SDSS J102402.60+062943.9 | 5 | Y | 13147.3 | 328.4 | 365.1 | AS | 1.074 | 9.99 | 11.04 | −22.26 | 2,8 |
| SDSS J102925.73+140823.2 | 5 | Yd | 18385.5 | 202.4 | 97.5 | AS | 0.120 | 9.32 | 10.11 | −17.97 | |
| SDSS J104326.47+110524.3 | 5 | Yd | 14285.1 | 338.0 | 209.5: | AS | 0.497 | 9.72 | 10.44: | −20.46: | 2 |
| SDSS J104913.79+044039.9 | 5 | Yd | ... | ... | ... | ... | <0.31 | <10.08 | ... | ... | 31 |
| SDSS J105115.42+054824.6 | ... | ... | ... | ... | ... | ... | ... | ... | ... | ... | 9,10 |
| SDSS J110538.99+020257.4 | ... | ... | ... | ... | ... | ... | ... | ... | ... | ... | 25 |
| SDSS J110640.20+051905.6 | 5 | Yd | 27488.5 | 331.0 | 212.1 | D | 0.309 | 10.11 | 10.92 | −20.50 | 2 |
| SDSS J110654.46+061213.1 | 5 | Yd | 13039.2 | 344.4 | 270.4: | D | 0.231 | 9.31 | 10.73: | −21.29: | |
| SDSS J111031.61+022043.2 | 5 | Y | 23732.3 | 385.5 | 267.0: | A | 1.311 | 10.61 | 10.60: | −21.24: | 2 |
| SDSS J111045.96+113641.6 | 5 | Y | 12698.3 | 257.7 | 170.6 | D | 2.034 | 10.23 | 10.67 | −19.79 | |
| SDSS J111237.43+120729.5 | 5 | Yd | 13989.1 | 480.1 | 232.7 | AS | 0.386 | 9.59 | 10.88 | −20.80 | 2,14 |
| SDSS J111639.15+040427.6 | 5 | Y | 22391.1 | 92.4 | 44.9: | S | 0.265 | 9.85 | 9.23: | −15.45: | 11 |
| SDSS J112328.11+052823.2 | ... | ... | ... | ... | ... | ... | ... | ... | ... | ... | 9 |
| SDSS J112455.86+080615.3 | 5 | Yd | 21091.6 | 403.8 | 257.1 | A | 0.168 | 9.60 | 10.83 | −21.12 | 2,32 |
| SDSS J112555.85+073107.8 | 5 | Yd | ... | ... | ... | ... | <0.26 | <9.84 | ... | ... | |
| SDSS J112813.02+102308.3 | 5 | Yd | ... | ... | ... | ... | <0.52 | <9.79 | ... | ... | 10 |
| SDSS J112904.01+061140.0 | 5 | Yd | ... | ... | ... | ... | <0.31 | <9.91 | ... | ... | |
| SDSS J113111.94+100231.3 | 5 | Y | 22330.0 | 468.3 | 219.9 | A | 0.718 | 10.29 | 10.84 | −20.61 | |
| SDSS J113249.28+101747.3 | 5 | Yd | 13373.7 | 105.9 | 37.0 | ? | 0.099 | 8.94 | 9.18 | −14.82 | 12 |
| SDSS J114008.71+030711.3 | 3 | Yd | 24427.1 | 328.8 | 661.7 | A | 0.476 | 10.19 | 11.61 | −24.19 | |
| SDSS J114105.71+024117.0 | ... | ... | ... | ... | ... | ... | ... | ... | ... | ... | 9 |

TABLE 2: H I Properties—*Continued*

| Source Name | Base. | Smooth | $v_{\rm sys}$ (km s$^{-1}$) | $W_{20}$ (km s$^{-1}$) | $v_m$ (km s$^{-1}$) | Prof. | $\int S_\nu dv$ (Jy km s$^{-1}$) | log $M_{\rm H\,I}$ ($M_\odot$) | log $M_{\rm dyn}$ ($M_\odot$) | $M_{B,{\rm TF}}$ (mag) | Notes |
|---|---|---|---|---|---|---|---|---|---|---|---|
| (1) | (2) | (3) | (4) | (5) | (6) | (7) | (8) | (9) | (10) | (11) | (12) |
| SDSS J115038.86+020854.2 | 5 | Yd | 32894.4 | 738.7 | 320.4 | D | 0.270 | 10.22 | 11.39 | −21.84 | 13 |
| SDSS J120011.27+100135.6 | 5 | Yd | 25642.7 | 208.0 | 196.4: | AS | 0.280 | 10.00 | 10.61: | −20.25: | 2 |
| SDSS J120257.81+045045.0 | 5 | N | 6273.7 | 431.4 | 205.8 | D | 4.628 | 9.95 | 10.87 | −20.40 | |
| SDSS J120330.23+123140.9 | 5 | Yd | 19308.4 | 327.9 | 170.3 | D | 0.502 | 10.00 | 10.58 | −19.78 | |
| SDSS J120332.94+022934.5 | 5 | N | 23214.2 | 394.5 | 225.6 | D | 1.062 | 10.49 | 11.05 | −20.70 | |
| SDSS J120648.31+065912.2 | 5 | Yd | ... | ... | ... | ... | <0.30 | <9.98 | ... | ... | |
| SDSS J121629.91+084253.3 | 5 | Yd | ... | ... | ... | ... | <0.25 | <9.78 | ... | ... | |
| SDSS J121930.87+064334.4 | 5 | Yd | 24112.3 | 189.4 | 250.6: | AS | 0.276 | 9.94 | 10.77: | −21.04: | |
| SDSS J122042.00+112405.2 | ... | ... | ... | ... | ... | ... | ... | ... | ... | ... | 33 |
| SDSS J122324.13+024044.4 | 5 | Yd | ... | ... | ... | ... | <0.50 | <9.10 | ... | ... | |
| SDSS J122811.42+095126.8 | 5 | Yd | ... | ... | ... | ... | <0.50 | <9.99 | ... | ... | 9 |
| SDSS J123235.82+060310.0 | 5 | Yd | ... | ... | ... | ... | <0.37 | <10.11 | ... | ... | |
| SDSS J123639.78+040758.4 | 5 | Y | 21272.9 | 593.5 | 261.5 | D | 0.820 | 10.29 | 11.17 | −21.18 | |
| SDSS J123944.23+120159.9 | 5 | Yd | 25664.6 | 275.4 | 125.8 | D | 0.220 | 9.90 | 10.44 | −18.80 | 2 |
| SDSS J124239.13+054717.6 | 5 | Y | 28135.8 | 248.8 | 171.2: | D | 0.326 | 10.15 | 10.50: | −19.80: | |
| SDSS J124319.98+025256.2 | 5 | Yd | ... | ... | ... | ... | <0.24 | <9.94 | ... | ... | |
| SDSS J124635.24+022208.7 | 5 | Yd | ... | ... | ... | ... | <0.47 | <9.70 | ... | ... | |
| SDSS J124913.75+151510.5 | ... | ... | ... | ... | ... | ... | ... | ... | ... | ... | 33 |
| SDSS J125039.08+141241.0 | 5 | Yd | ... | ... | ... | ... | <0.58 | <10.31 | ... | ... | |
| SDSS J130241.53+040738.6 | 5 | Yd | ... | ... | ... | ... | <0.34 | <10.25 | ... | ... | 9 |
| SDSS J130803.04+035114.5 | 5 | Yd | ... | ... | ... | ... | <0.34 | <9.91 | ... | ... | |
| SDSS J132026.49+051113.5 | 5 | Yd | 29483.1 | 454.3 | 236.7: | D | 0.229 | 10.05 | 10.75: | −20.85: | 9 |
| SDSS J132442.44+052438.8 | ... | ... | ... | ... | ... | ... | ... | ... | ... | ... | 9 |
| SDSS J134952.84+020445.1 | 5 | Y | 9831.1 | 355.1 | 172.9 | D | 1.172 | 9.76 | 10.48 | −19.83 | 2,9 |
| SDSS J140018.42+050242.2 | 5 | Y | 10322.6 | 364.9 | 234.1 | D | 0.602 | 9.51 | 10.72 | −20.82 | 2 |
| SDSS J142748.28+050222.0 | 5 | Y | 31763.8 | 167.6 | 203.7: | D | 0.251 | 10.16 | 10.51: | −20.37: | 9 |
| SDSS J143450.62+033842.5 | 5 | Yd | ... | ... | ... | ... | <0.87 | <9.50 | ... | ... | 15 |
| SDSS J145047.19+033645.4 | 5 | Yd | ... | ... | ... | ... | <0.55 | <10.07 | ... | ... | 9,10 |
| SDSS J150556.55+034226.3 | 5 | Y | 10822.0 | 443.4 | 222.8 | D | 0.892 | 9.72 | 10.94 | −20.66 | 2,9 |
| SDSS J153839.19+024324.7 | 5 | Yd | 9497.8 | 428.2 | 253.3 | D | 1.526 | 9.84 | 10.78 | −21.07 | |
| SDSS J155417.43+323837.8 | 5 | Yd | 14494.2 | 311.3 | 225.8 | AS | 0.350 | 9.58 | 10.55 | −20.70 | 2 |
| SDSS J160936.42+254459.6 | 5 | Yd | 12476.7 | 436.5 | 398.6: | D | 0.631 | 9.69 | 10.86: | −22.55: | 2 |
| SDSS J163152.23+345328.7 | 5 | Yd | 21702.5 | 241.2 | 173.9 | A? | 0.276 | 9.86 | 10.59 | −19.85 | 29 |
| SDSS J163159.59+243740.3 | 5 | Yd | ... | ... | ... | ... | <0.30 | <9.42 | ... | ... | |
| SDSS J163453.67+231242.7 | 5 | Yd | 11593.0 | 297.0 | 195.0 | D | 0.456 | 9.49 | 10.55 | −20.22 | |
| SDSS J163523.44+263046.5 | 5 | Yd | ... | ... | ... | ... | <0.34 | <9.91 | ... | ... | |
| SDSS J165601.61+211241.2 | 5 | Yd | 14715.2 | 319.1 | 150.2: | A | 0.361 | 9.61 | 10.20: | −19.38: | 29 |
| SDSS J170102.29+340400.5 | 5 | Yd | ... | ... | ... | ... | <0.28 | <10.10 | ... | ... | |
| SDSS J171257.02+231526.8 | 5 | Yd | ... | ... | ... | ... | <0.28 | <9.64 | ... | ... | |
| SDSS J171520.19+274825.9 | 5 | Yd | ... | ... | ... | ... | <0.32 | <10.27 | ... | ... | |
| SDSS J171601.93+311213.8 | ... | ... | ... | ... | ... | ... | ... | ... | ... | ... | 9 |
| SDSS J210622.99+105606.5 | 5 | Yd | 28420.6 | 276.5 | 174.4 | AS | 0.163 | 9.86 | 10.46 | −19.86 | 2 |
| SDSS J210721.91+110359.0 | 5 | Yd | ... | ... | ... | ... | <0.62 | <9.71 | ... | ... | 9 |
| SDSS J211646.34+110237.4 | 5 | Yd | 24186.9 | 322.5 | 322.2 | AS | 0.304 | 9.98 | 11.19 | −21.85 | |
| SDSS J223338.42+131243.6 | 5 | Yd | 28165.8 | 550.1 | 263.9 | D | 0.312 | 10.13 | 11.15 | −21.21 | 2 |
| SDSS J225305.11+140351.2 | 5 | Yd | ... | ... | ... | ... | <0.32 | <10.13 | ... | ... | |
| SDSS J232721.96+152437.3 | ... | ... | ... | ... | ... | ... | ... | ... | ... | ... | 9,16 |
| SDSS J235128.78+155259.0 | 5 | Yd | ... | ... | ... | ... | <0.48 | <10.35 | ... | ... | |
| 3C 120 | ... | ... | ... | ... | ... | ... | ... | ... | ... | ... | 16 |
| Akn 120 | 5 | Y | 9809.2 | 370.3 | 266.1 | D | 1.965 | 9.98 | 11.28 | −21.23 | |

TABLE 2: H I Properties—*Continued*

| Source Name (1) | Base. (2) | Smooth (3) | $v_{\rm sys}$ (km s$^{-1}$) (4) | $W_{20}$ (km s$^{-1}$) (5) | $v_m$ (km s$^{-1}$) (6) | Prof. (7) | $\int S_\nu dv$ (Jy km s$^{-1}$) (8) | log $M_{\rm H\,I}$ ($M_\odot$) (9) | log $M_{\rm dyn}$ ($M_\odot$) (10) | $M_{B,\rm TF}$ (mag) (11) | Notes (12) |
|---|---|---|---|---|---|---|---|---|---|---|---|
| MCG +01-13-012 | 5 | N | 4705.1 | 440.7 | 216.2 | D | 3.150 | 9.52 | 10.99 | −20.56 | 17 |
| NGC 3227 | 5 | N | 1135.6 | 453.4 | 235.5 | D | 15.495 | 9.03 | 10.89 | −20.84 | 34 |
| NGC 5548 | 5 | Y | 5169.8 | 321.1 | 220.8 | AS | 1.384 | 9.26 | 10.99 | −20.63 | 1,17 |
| NGC 7469 | 7 | Y | 4899.5 | 525.1 | 414.4 | AS | 3.741 | 9.65 | 11.53 | −22.67 | 2,18,35 |
| PG 0003+199 | 5 | Yd | ⋯ | ⋯ | ⋯ | ⋯ | <0.35 | <9.02 | ⋯ | ⋯ | 19 |
| PG 0844+349 | 5 | Yd | 19363.1 | 488.5 | 213.9L | D | 1.048 | 10.31 | ⋯ | −20.52L | 2,20 |
| PG 1211+143 | ⋯ | ⋯ | ⋯ | ⋯ | ⋯ | ⋯ | ⋯ | ⋯ | ⋯ | ⋯ | 16 |
| PG 1229+204 | 5 | Yd | 19208.2 | 295.0 | 152.7 | ? | 0.229 | 9.64 | 10.12 | −19.43 | |
| PG 1307+085 | ⋯ | ⋯ | ⋯ | ⋯ | ⋯ | ⋯ | ⋯ | ⋯ | ⋯ | ⋯ | 16 |
| PG 1426+015 | 5 | Yd | 25975.2 | 357.4 | 148.9L | D | 0.456 | 10.23 | 10.11 | −19.35L | |
| PG 2130+099 | 5 | Yd | 19000.5 | 506.0 | 275.8 | D | 0.469 | 9.95 | 10.87 | −21.35 | 1 |
| RX J0602.1+2828 | 5 | Y | 9907.0 | 627.2 | 287.7L | AS | 2.111 | 10.02 | ⋯ | −21.49L | 9 |
| RX J0608.0+3058 | 5 | Y | 21869.1 | 218.9 | 86.3L | D | 0.175 | 9.65 | ⋯ | −17.58L | |

NOTE.— Col. (1) Source name. Col. (2) Order of polynomial used to fit the baseline. Col. (3) Type of smoothing applied to the data prior to fitting. Col. (4) Systemic velocity in barycentric optical frame. Col. (5) Width of the line profile measured at 20% of the peak. Col. (6) Maximum rotational velocity, corrected for inclination, instrumental resolution, and redshift; derived from $W_{20}$. Entries followed by a colon are particularly uncertain because of the uncertain inclination correction; those followed by "L" denote lower limits because inclination angles are not available (see Table 1, Col. 6). Col. (7) Profile type. "D" = classical double-peaked; "A" = asymmetric; "S" = single-peaked; "?" = signal-to-noise ratio too low to tell. Col. (8) Integrated line flux; upper limits are calculated as the product of 3 times the root-mean-square noise level and a rest-frame line width of 304 km s$^{-1}$ (see text). Col. (9) H I mass, derived from Equation 1. Col. (10) Dynamical mass of the galaxy, derived from $D_{25}$ and $v_m$ using Equation 4. Col. (11) $B$-band absolute magnitude of the host galaxy, derived from $v_m$ and the Tully-Fisher relation (see text). Col. (12) Notes: (1) Faint neighboring galaxy in the beam, but unlikely to be a source of confusion. (2) Possible confusion. (3) RFI at high side of line; width might be affected at the level of ∼ 20 km s$^{-1}$. (4) Uncertain baseline. (5) 3.4 $\sigma$ feature; unclear if real. (6) Low S/N. (7) Low level broad line might be present. (8) Galaxy pair. (9) Solar RFI. (10) Close to 1350 MHz radar. (11) Nearby RFI does not affect H I. (12) Dubious detection; possible H I within redshift uncertainty. (13) Uncertain; RFI at H I edge, if H I is real. (14) Very broad profile. (15) RMS includes some RFI. (16) Standing waves (strong continuum). (17) Feed res.; < line power. (18) Feed res.; used YY polarization only. (19) 3.3 $\sigma$ dip at ∼ 7885 km s$^{-1}$; unlikely to be real. (20) Close apparent interacting companion in *HST* image (Fig. 3*h*). (21) Integration time too short. (22) Difficult profile to measure. (23) Peak is 2.4 $\sigma$. (24) Narrow line; much bandpass structure. (25) Peak is < 3 $\sigma$. (26) Irregular profile, but strong peak. (27) Peak at 2.8 $\sigma$. (28) Odd dip on low-velocity side of line; Gaussian fit to feature shows negligible impact on H I profile and properties. (29) Possible RFI contamination. (30) Feature at 18150 km s$^{-1}$ is RFI; broad shoulders. (31) High feature at 3.4 $\sigma$. (32) Possible H I; line width very sensitive to choice of initial parameters of fit. (33) Bad bandpasses. (34) Interacting with NGC 3226, but the latter is unlikely to be a source of confusion because it is an elliptical. (35) Interacting with IC 5283, an Scd galaxy, so likely to be a source of confusion.

TABLE 3: Comparison with Published Work

| Source Name | $W_{20}$ (km s$^{-1}$) | | $\int S_\nu dv$ (Jy km s$^{-1}$) | |
|---|---|---|---|---|
| | Ours | Literature | Ours | Literature |
| IC 492 | 263.8 | 257.0±3.9 | 2.93 | 3.44±0.61 |
| IC 756 | 431.4 | 402.7−444.6 | 4.63 | 4.83±0.85 |
| MCG +01-13-012 | 440.7 | 442.6±12.1 | 3.15 | 2.68±0.22 |
| Mrk 1146 | 407.9 | 417.8±16.6 | 1.61 | 2.83±1.13 |
| NGC 3227 | 453.4 | 299.9−526.0 | 15.49 | 18.88±2.44 |
| NGC 5548 | 321.1 | 239.9−297.2 | 1.38 | 1.38±0.17 |
| NGC 7469 | 525.1 | 395.4−510.5 | 3.74 | 2.73±0.61 |
| PG 2130+099 | 506.0 | 950.6±79.6 | 0.47 | 0.97±0.39 |

NOTE.—Literature values compiled from Hyperleda (`http://leda.univ-lyon1.fr`).

TABLE 4: Spectroscopic Properties of SDSS Objects

| Source Name | [O II] λ3727 | Hβ λ4861 | [O III] λ5007 | Hα λ6563 | [N II] λ6583 | [S II] λ6716 | [S II] λ6731 | $M_{B,\mathrm{host}}$ (mag) |
|---|---|---|---|---|---|---|---|---|
| (1) | (2) | (3) | (4) | (5) | (6) | (7) | (8) | (9) |
| SDSS J000805.62+145023.3 | 0.191 | 0.208 | 1.89 | 1.034 | 0.142 | 0.185 | 0.197 | −20.39 |
| SDSS J003646.45+145936.9 | 0.038 | 0.082 | 4.12 | 0.253 | 0.194 | 0.085 | 0.060 | −20.20 |
| SDSS J004055.88+153349.0 | 0.105 | 0.213 | 3.60 | 0.872 | 0.480 | 0.179 | 0.167 | −20.99 |
| SDSS J004719.39+144212.6 | 0.019 | 0.079 | 20.47 | 0.248 | 0.227 | 0.063 | 0.049 | −21.46 |
| SDSS J010712.03+140844.9 | 0.033 | 0.061 | 3.49 | 0.190 | 0.104 | 0.045 | 0.045 | > −19.42 |
| SDSS J015046.68+132359.9 | 0.218 | 0.463 | 195.34 | 1.156 | 0.161 | 0.068 | 0.082 | −19.11 |
| SDSS J075245.60+261735.7 | 0.049 | 0.111 | 7.99 | 0.345 | 0.271 | 0.072 | 0.086 | −20.40 |
| SDSS J080243.39+310403.3 | 0.225 | 0.105 | 1847.99 | 0.295 | 0.293 | 0.076 | 0.077 | −19.92 |
| SDSS J080538.66+261005.4 | ⋯ | 0.066 | 27.45 | 0.303 | 0.312 | 0.083 | 0.078 | −20.17 |
| SDSS J080546.97+260532.9 | 0.112 | 0.049 | 19.42 | 0.441 | 0.396 | 0.143 | 0.133 | −20.39 |
| SDSS J081700.40+343556.3 | 2.070 | 0.385 | 2064.29 | 1.188 | 0.660 | 0.480 | 0.391 | −19.34 |
| SDSS J081835.73+285022.4 | 0.022 | 0.027 | 3.42 | 0.085 | <0.015 | 0.064 | 0.045 | −20.12 |
| SDSS J082048.30+282217.8 | 0.346 | 0.097 | 246.49 | 0.508 | 0.453 | 0.138 | 0.111 | −19.58 |
| SDSS J082320.68+074020.2 | 0.366 | 0.076 | 123.28 | 0.223 | 0.600 | 0.152 | 0.105 | −20.88 |
| SDSS J083045.36+340532.1 | 0.109 | 0.093 | 4509.09 | 0.333 | 0.253 | 0.069 | 0.061 | −20.06 |
| SDSS J083107.63+052105.9 | 0.365 | 0.122 | 1122.54 | 0.581 | 0.404 | 0.167 | 0.140 | −19.32 |
| SDSS J083747.89+053644.8 | 0.343 | 0.096 | 267.48 | 0.415 | 0.424 | 0.195 | 0.170 | −19.51 |
| SDSS J084025.51+033301.7 | 0.036 | 0.763 | 1454.20 | 0.070 | 0.070 | 0.025 | 0.026 | −19.66 |
| SDSS J084556.67+340936.2 | 0.080 | 0.079 | 988.46 | 0.222 | 0.258 | 0.040 | 0.040 | −20.91 |
| SDSS J090028.53+060842.5 | 0.054 | 0.167 | 2.18 | 0.811 | 0.401 | 0.156 | 0.092 | −19.32 |
| SDSS J091222.31+065829.0 | 0.286 | 0.285 | 2.87 | 0.884 | 0.626 | 0.198 | 0.210 | −18.74 |
| SDSS J093240.55+023332.6 | 0.195 | 0.107 | 1049.79 | 0.338 | 0.227 | 0.076 | 0.064 | −20.09 |
| SDSS J093259.60+040506.0 | 0.353 | 0.188 | 1020.02 | 0.762 | 0.476 | 0.188 | 0.151 | −20.67 |
| SDSS J093812.26+074340.0 | 0.000 | 0.135 | 1427.78 | 0.477 | 0.662 | 0.015 | 0.044 | −19.36 |
| SDSS J093917.26+363343.9 | 0.000 | 0.318 | 1534.84 | 1.229 | 0.435 | 0.688 | 0.585 | −19.42 |
| SDSS J094529.37+093610.4 | 0.000 | 0.069 | 9602.33 | 0.276 | 0.197 | 0.062 | 0.061 | −18.20 |
| SDSS J095955.85+130237.8 | 0.099 | 0.164 | 1153.99 | 0.444 | 0.246 | 0.064 | 0.066 | −21.10 |
| SDSS J100155.79+055413.3 | 0.356 | 0.094 | 428.27 | 0.549 | 0.521 | 0.225 | 0.169 | −19.91 |
| SDSS J102148.90+030732.2 | 0.265 | 0.172 | 1183.84 | 0.559 | 0.343 | 0.152 | 0.115 | −20.70 |
| SDSS J102402.60+062943.9 | 0.044 | 0.069 | 7.10 | 0.216 | 0.148 | 0.071 | 0.061 | −20.11 |
| SDSS J102925.73+140823.2 | 0.206 | 0.137 | 1112.72 | 0.674 | 0.541 | 0.128 | 0.113 | −21.03 |
| SDSS J104326.47+110524.3 | 0.336 | 0.129 | 7089.73 | 0.399 | 0.043 | 0.050 | 0.043 | −17.13 |
| SDSS J104913.79+044039.9 | 1.997 | 0.753 | 122.61 | 3.262 | 1.407 | 0.750 | 0.532 | −19.86 |
| SDSS J105115.42+054824.6 | 0.084 | 0.148 | 8.69 | 0.573 | 0.423 | 0.145 | 0.129 | −19.22 |
| SDSS J110538.99+020257.4 | 0.032 | 0.045 | 1633.48 | 0.124 | 0.108 | 0.027 | 0.033 | −21.49 |
| SDSS J110640.20+051905.6 | 0.169 | 0.064 | 439.41 | 0.244 | 0.398 | 0.096 | 0.077 | −21.32 |
| SDSS J110654.46+061213.1 | 0.321 | 0.265 | 2.60 | 0.822 | 0.381 | 0.565 | 0.468 | −19.22 |
| SDSS J111031.61+022043.2 | 0.151 | 0.199 | 2.04 | 0.989 | 0.136 | 0.177 | 0.188 | −19.35 |
| SDSS J111045.96+113641.6 | 0.178 | 0.112 | 718.17 | 0.456 | 0.438 | 0.142 | 0.141 | −20.98 |
| SDSS J111237.43+120729.5 | 0.258 | 0.206 | 358.23 | 1.750 | 1.336 | 0.328 | 0.229 | −20.23 |
| SDSS J111639.15+040427.6 | 0.065 | 0.093 | 7.51 | 0.454 | 0.358 | 0.119 | 0.120 | −19.58 |
| SDSS J112328.11+052823.2 | 0.205 | 0.122 | 614.93 | 0.492 | 0.276 | 0.084 | 0.059 | −19.79 |
| SDSS J112455.86+080615.3 | 0.415 | 0.455 | 432.02 | 2.014 | 0.990 | 0.352 | 0.258 | −21.20 |
| SDSS J112555.85+073107.8 | 0.218 | 0.100 | 391.24 | 0.292 | 0.226 | 0.082 | 0.063 | −20.47 |
| SDSS J112813.02+102308.3 | 0.048 | 0.144 | 30.67 | 0.431 | 0.192 | 0.055 | 0.049 | −19.89 |
| SDSS J112904.01+061140.0 | 0.146 | 0.086 | 2992.35 | 0.213 | 0.103 | 0.038 | 0.044 | −19.61 |
| SDSS J113111.94+100231.3 | 0.049 | 0.215 | 2.54 | 0.958 | 0.572 | 0.198 | 0.142 | −20.54 |
| SDSS J113249.28+101747.3 | 0.022 | 0.100 | 22.53 | 0.363 | 0.458 | 0.094 | 0.081 | −20.09 |
| SDSS J114008.71+030711.3 | ⋯ | 0.639 | 2.07 | 1.981 | 1.239 | 0.284 | 0.285 | −20.07 |
| SDSS J114105.71+024117.0 | 0.185 | 0.132 | 757.66 | 0.382 | 0.224 | 0.071 | 0.064 | −20.20 |

TABLE 4: Spectroscopic Properties of SDSS Objects—*Continued*

| Source Name (1) | [O II] λ3727 (2) | Hβ λ4861 (3) | [O III] λ5007 (4) | Hα λ6563 (5) | [N II] λ6583 (6) | [S II] λ6716 (7) | [S II] λ6731 (8) | $M_{B,\mathrm{host}}$ (mag) (9) |
|---|---|---|---|---|---|---|---|---|
| SDSS J115038.86+020854.2 | 0.038 | 0.043 | 2.23 | 0.593 | 0.665 | 0.209 | 0.201 | −20.26 |
| SDSS J120011.27+100135.6 | 0.065 | 0.067 | 31.72 | 0.209 | 0.173 | 0.042 | 0.040 | −20.71 |
| SDSS J120257.81+045045.0 | ⋯ | 0.089 | 14.28 | 0.476 | 0.698 | 0.183 | 0.198 | −20.29 |
| SDSS J120330.23+123140.9 | 0.297 | 0.298 | 86.47 | 1.482 | 0.849 | 0.236 | 0.175 | −20.50 |
| SDSS J120332.94+022934.5 | 0.083 | 0.101 | 29.17 | 0.380 | 0.283 | 0.080 | 0.075 | −21.61 |
| SDSS J120648.31+065912.2 | 0.216 | 0.072 | 1564.87 | 0.271 | 0.265 | 0.092 | 0.089 | −20.01 |
| SDSS J121629.91+084253.3 | 0.233 | 0.089 | 1152.96 | 0.317 | 0.472 | 0.195 | 0.179 | −19.27 |
| SDSS J121930.87+064334.4 | 0.266 | 0.243 | 561.79 | 0.812 | 0.399 | 0.145 | 0.115 | −20.89 |
| SDSS J122042.00+112405.2 | 0.204 | 0.143 | 588.90 | 0.497 | 0.295 | 0.086 | 0.090 | −20.94 |
| SDSS J122324.13+024044.4 | ⋯ | 0.121 | 16.95 | 0.443 | 0.231 | 0.092 | 0.079 | −18.57 |
| SDSS J122811.42+095126.8 | 0.098 | 0.079 | 8.79 | 0.352 | 0.614 | 0.226 | 0.191 | −20.25 |
| SDSS J123235.82+060310.0 | 0.140 | 0.114 | 5.41 | 0.568 | 0.296 | 0.127 | 0.098 | −20.42 |
| SDSS J123639.78+040758.4 | 0.103 | 0.069 | 7.55 | 0.400 | 0.698 | 0.193 | 0.210 | −20.51 |
| SDSS J123944.23+120159.9 | 0.952 | 0.175 | 72.42 | 1.435 | 1.246 | 0.512 | 0.445 | −21.25 |
| SDSS J124239.13+054717.6 | 0.061 | 0.105 | 5.92 | 0.524 | 0.273 | 0.117 | 0.090 | −20.15 |
| SDSS J124319.98+025256.2 | 0.206 | 0.155 | 727.38 | 0.443 | 0.290 | 0.073 | 0.076 | −20.17 |
| SDSS J124635.24+022208.7 | 0.123 | 0.310 | 24.58 | 0.961 | 0.352 | 0.101 | 0.101 | −20.09 |
| SDSS J124913.75+151510.5 | 0.594 | 0.226 | 102.85 | 1.939 | 1.904 | 0.688 | 0.478 | −21.13 |
| SDSS J125039.08+141241.0 | 0.618 | 0.189 | 1524.23 | 0.950 | 0.552 | 0.267 | 0.231 | −20.89 |
| SDSS J130241.53+040738.6 | 0.151 | 0.073 | 3.98 | 0.398 | 0.154 | 0.103 | 0.076 | −19.48 |
| SDSS J130803.04+035114.5 | 0.032 | 0.079 | 17.76 | 0.244 | 0.293 | 0.085 | 0.078 | −20.06 |
| SDSS J132026.49+051113.5 | 0.053 | 0.119 | 6.54 | 0.385 | 0.245 | 0.126 | 0.099 | −20.10 |
| SDSS J132442.44+052438.8 | 0.420 | 0.239 | 1.75 | 1.194 | 0.164 | 0.213 | 0.227 | −18.08 |
| SDSS J134952.84+020445.1 | 0.128 | 0.107 | 124.03 | 0.349 | 0.187 | 0.077 | 0.077 | −19.38 |
| SDSS J140018.42+050242.2 | 6.752 | 0.215 | 1.81 | 1.069 | 0.147 | 0.191 | 0.204 | −20.77 |
| SDSS J142748.28+050222.0 | 0.018 | 0.134 | 12.64 | 4.527 | <0.012 | 0.016 | 0.033 | −21.55 |
| SDSS J143450.62+033842.5 | ⋯ | 0.236 | 5.29 | 0.823 | 0.494 | 0.160 | 0.147 | −18.72 |
| SDSS J145047.19+033645.4 | 0.305 | 0.244 | 1.33 | 0.757 | 0.642 | 0.212 | 0.178 | −18.57 |
| SDSS J150556.55+034226.3 | 0.052 | 0.080 | 231.96 | 0.302 | 0.329 | 0.080 | 0.086 | −21.41 |
| SDSS J153839.19+024324.7 | 4.964 | 1.551 | 24.84 | 5.477 | 1.308 | 1.577 | 1.516 | −18.39 |
| SDSS J155417.43+323837.8 | 0.219 | 0.115 | 5707.56 | 0.500 | 0.346 | 0.133 | 0.115 | −19.96 |
| SDSS J160936.42+254459.6 | 0.574 | 0.190 | 1689.61 | 0.530 | 0.176 | 0.086 | 0.075 | −19.15 |
| SDSS J163152.23+345328.7 | 0.172 | 0.056 | 433.73 | 0.207 | 0.408 | 0.134 | 0.089 | −20.94 |
| SDSS J163159.59+243740.3 | 0.378 | 0.075 | 1078.42 | 0.297 | 0.161 | 0.098 | 0.075 | −18.75 |
| SDSS J163453.67+231242.7 | 0.281 | 0.113 | 2257.33 | 0.489 | 0.451 | 0.179 | 0.140 | −21.02 |
| SDSS J163523.44+263046.5 | 0.141 | 0.082 | 447.23 | 0.293 | 0.307 | 0.099 | 0.089 | −20.36 |
| SDSS J165601.61+211241.2 | 0.360 | 0.165 | 3935.65 | 0.625 | 0.162 | 0.112 | 0.084 | −19.35 |
| SDSS J170102.29+340400.5 | 0.216 | 0.096 | 2841.40 | 0.333 | 0.145 | 0.048 | 0.051 | > −19.69 |
| SDSS J171257.02+231526.8 | 1.222 | 0.332 | 65.60 | 1.475 | 1.876 | 0.924 | 0.772 | −20.13 |
| SDSS J171520.19+274825.9 | 0.383 | 0.098 | 513.18 | 0.320 | 0.301 | 0.180 | 0.154 | −19.72 |
| SDSS J171601.93+311213.8 | 0.156 | 0.062 | 7727.33 | 0.168 | 0.199 | 0.056 | 0.050 | −21.57 |
| SDSS J210622.99+105606.5 | 0.208 | 0.236 | 4.99 | 0.818 | 0.689 | 0.296 | 0.212 | −20.63 |
| SDSS J210721.91+110359.0 | 0.109 | 0.096 | 6.17 | 0.479 | 0.249 | 0.107 | 0.082 | −19.26 |
| SDSS J211646.34+110237.4 | 0.092 | 0.026 | 85.07 | 0.180 | 0.086 | 0.084 | 0.073 | −21.28 |
| SDSS J223338.42+131243.6 | 0.065 | 0.073 | 23.98 | 0.227 | 0.210 | 0.089 | 0.079 | −21.58 |
| SDSS J225305.11+140351.2 | 0.013 | 0.024 | 92.95 | 0.167 | 0.079 | 0.077 | 0.067 | −20.33 |
| SDSS J232721.96+152437.3 | 0.018 | 0.022 | 97.22 | 0.153 | 0.073 | 0.071 | 0.062 | −21.28 |
| SDSS J235128.78+155259.0 | 0.329 | 0.113 | 2991.89 | 0.455 | 0.357 | 0.125 | 0.120 | −20.56 |

NOTE.— Col. (1) Source name. Col. (2)–(8) All fluxes are relative to that of [O III] λ5007, which is given in units of $10^{-15}$ ergs s$^{-1}$ cm$^{-2}$. Galactic extinction has been corrected, but not internal extinction. Col. (9) $B$-band absolute magnitude of the host galaxy, derived from subtracting the AGN contribution from the total luminosity (see text). In the case of the SDSS objects, the $g$-band magnitudes were converted to the $B$ band assuming $g - B = -0.45$ mag, appropriate for an Sab galaxy (Fukugita et al. 1995).

$$v_m = \frac{(W_{20} - W_{\rm inst})/(1+z) - W_{\rm turb}}{2\sin i}. \quad (1)$$

In the optically thin limit, the integrated line flux, $\int S_\nu\, dv$, in units of Jy km s$^{-1}$, is related to the H I mass as (Roberts 1962)

$$M_{\rm H\,I} = 2.36 \times 10^5\, D_L^2 \int S_\nu\, dv\ M_\odot, \quad (2)$$

where $D_L$ is the luminosity distance expressed in Mpc and $dv$ is the line width in the observer's frame. We neglect any correction for self-absorption, since this is controversial (see, e.g., Springob et al. 2005), and, in any case, depends on Hubble type, which is not well-known for many of our sources (see §2.3). Upper limits for the integrated fluxes and H I masses are calculated using 3 times the root-mean-square noise level and a rest-frame line width of 304 km s$^{-1}$, the median value for the 66 detected objects.

Single-dish H I observations always run the risk of source confusion, especially for relatively distant samples such as ours. At the median redshift of $z = 0.05$ for our targets, Arecibo's telescope beam (FWHM $\approx 3'.5$) subtends a linear diameter of $\sim 200$ kpc. We use the optical images (from SDSS if available, or else from the Palomar Digital Sky Survey; see §2.3), in combination with the redshifts, to identify potential sources of confusion within a search radius of $7'.5$. The intensity of the first sidelobes of the Arecibo beam drops to $\sim 10\%$ of the peak at a distance of $5'.5$ from the beam center, and by $7'$–$8'$ it becomes negligible (Heiles et al. 2000). We consider an object as a candidate confusing source if it lies within the search radius and has a cataloged radial velocity within $\pm 500$ km s$^{-1}$ of that of the science target. Only a few candidates have been identified, and these are noted in Table 2. The vast majority of the objects in our survey are unaffected by source confusion.

Eight of the objects in our survey have published H I data. A comparison of our measurements with those in the literature (Table 3) shows that in general there is fairly good agreement. The most noticeable exception is PG 2130+099, for which both our line width and flux are lower than the literature values by about a factor of 2.

### 2.3. *Optical Data*

We use both optical spectroscopic and imaging data to ascertain a number of parameters that are central to our analysis. For the SDSS objects, these data were taken directly from the SDSS archives. The spectra were analyzed following the procedures previously described in Greene & Ho (2004, 2005b; see also Kim et al. 2006). In brief, we obtain a pure emission-line spectrum for each object by subtracting from the observed total spectrum a model consisting of a stellar component, a featureless power-law component, and an Fe II "pseudo-continuum." We then fit the resulting narrow and broad emission lines using a combination of multi-component Gaussians. The optical emission-line parameters are collected in Table 4. We also give (in Table 1), where available, values of the central stellar velocity dispersion and its associated uncertainty, derived using the technique of Greene & Ho (2006a). If the data do not permit the stellar velocity dispersion to be measured, we list instead the velocity dispersion of the [O II] $\lambda 3727$ line, which Greene & Ho (2005a) have shown to be an effective substitute. BH masses were estimated using the broad H$\alpha$ method of Greene & Ho (2005b), using the FHWM and luminosities given in Table 1. We further convert the broad H$\alpha$ luminosity to the AGN continuum luminosity at 5100 Å, using Equation 1 of Greene & Ho (2005b), from which we deduce the bolometric luminosity assuming that $L_{\rm bol} = 9.8\, L_{5100}$ (McClure & Dunlop 2004).

The non-SDSS objects were treated differently. The majority of these, by design, have BH masses directly measured from reverberation mapping, and we simply adopt the values given in Peterson et al. (2004), from which continuum luminosities at 5100 Å were also taken. Three of the non-SDSS objects (MCG +01-13-012, RX J0602.1+2828, and RX J0608.0+3058) only have measurements for the H$\beta$ line, but BH masses based on this line alone can also be estimated with reasonable accuracy (Greene & Ho 2005b).

The images provide five important pieces of information about the sources: the total (AGN plus host galaxy) magnitude, morphological type, size, inclination angle, and potential sources of confusion within the H I beam. For the SDSS objects, we choose the $g$ band as our fiducial reference point, since it is closest to the more traditional $B$ band on which most of the literature references are based. In Figure 3, we display the optical image of the sources detected in H I; images of the H I nondetections are shown in Figure 4. In a few cases we were able to locate high-resolution images in the *Hubble Space Telescope (HST)* archives. The size of each image has been scaled to a constant physical scale of 50 kpc $\times$ 50 kpc to facilitate comparison of objects with very different distances.

Inspection of Figures 3 and 4 shows that obtaining reliable morphological types of the host galaxies is challenging for most of the sources, because of their small angular sizes and the coarse resolution and shallow depth of the SDSS images. In assigning a morphological type, we must be careful to give lower weight to the apparent prominence of the bulge, since a substantial fraction of the central brightness enhancement presumably comes from the AGN core itself. The SDSS database provides quantitative measurements of the Petrosian radius containing 50% and 90% of the light, from which one can calculate the (inverse) "concentration index," defined to be $C \equiv r_{\rm P50}/r_{\rm P90}$. We use the correlation between $C$ and morphological type index of Shimasaku et al. (2001) as an additional guide to help us assign morphological types, again bearing in mind that because of the AGN contamination the concentration index should be viewed strictly as an upper limit to the true value. We generally give less weight to the classifications based on $C$. (We have discovered a few glaring examples where the SDSS-based concentration index gives an egregiously erroneous morphological type.) The most difficult classifications are those that lie on the boundary between ellipticals and S0s, which is sometimes ambiguous even for nearby, bright galaxies. Unless the galaxy is highly inclined, it is often just impossible to tell; we label these cases as "E/S0." Another difficult situation arises when trying to discern whether a disk galaxy truly possesses spiral arms. Given the modest quality of the SDSS images and the relatively large distances of the galaxies, again often no clear-cut decision can be made, and we are forced to assign a classification of "S0/Sp." For a few of the objects, the image material is simply inadequate to allow a classification to be made at all.

The SDSS photometry additionally provides values for the major axis ($a$) and minor axis ($b$) isophotal diameters measured at a surface brightness level of $\mu = 25$ mag arcsec$^{-2}$, from which we can deduce the photometric inclination angle using Hubble's (1926) formula

$$\cos^2 i = \frac{q^2 - q_0^2}{1 - q_0^2}, \quad (3)$$

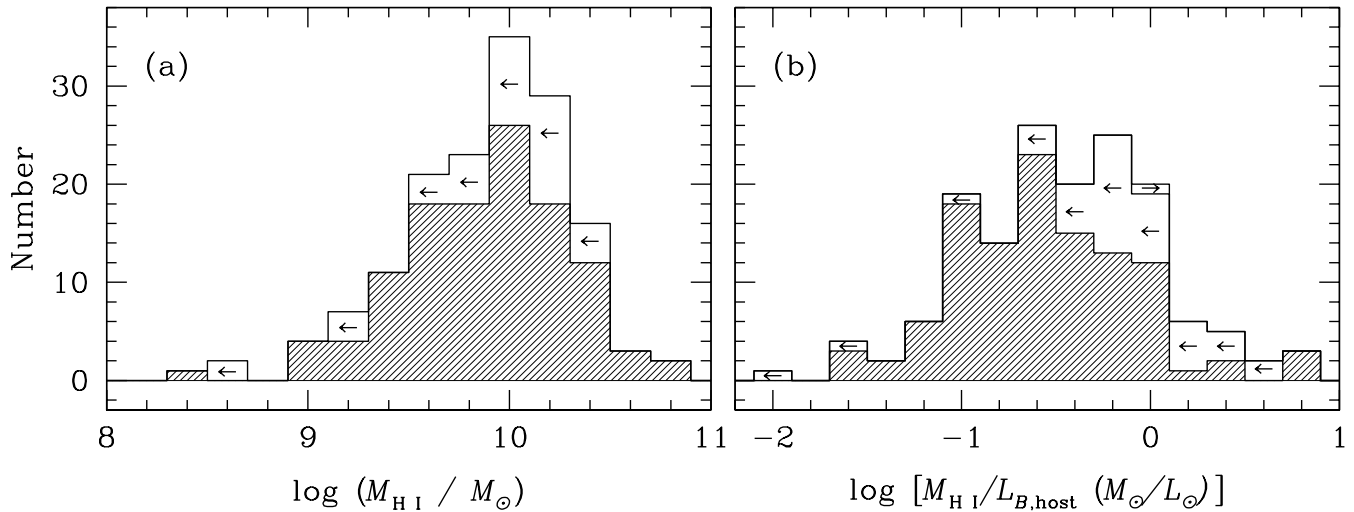

FIG. 5.— The distribution of (*a*) H I masses and (*b*) H I masses normalized to the *B*-band luminosity of the host galaxy. Limits are plotted as open histograms.

where $q = b/a$. The intrinsic thickness of the disk, $q_0$, varies by about a factor of 2 along the spiral sequence; we adopt $q_0 = 0.3$, a value appropriate for early-type systems (Fouqué et al. 1990). It is also of interest to combine the galaxy's optical size ($D_{25}$, diameter at $\mu = 25$ mag arcsec$^{-2}$) with the H I line width to compute a characteristic dynamical mass. From Casertano & Shostak (1980),

$$M_{\rm dyn} = 2 \times 10^4 \left(\frac{D_L}{\rm Mpc}\right) \left(\frac{D_{25}}{\rm arcmin}\right) \left(\frac{v_m}{\rm km\ s^{-1}}\right)^2\ M_\odot. \quad (4)$$

Because we have no actual measurement of the size of the H I disk, this formula yields only an approximate estimate of the true dynamical mass. However, from spatially resolved observations we know that the sizes of H I disks of spiral galaxies, over a wide range of Hubble types and luminosities, scale remarkably well with their optical sizes. From the studies of Broeils & Rhee (1997) and Noordermeer et al. (2005), $D_{\rm H I}/D_{25} \approx 1.7$ within 30%–40%. Nevertheless, our values of $M_{\rm dyn}$ are probably much more accurate as a relative rather than an absolute measure of the galaxy dynamical mass.

The optical photometry, albeit of insufficient angular resolution to yield a direct decomposition of the host galaxy from the AGN core, nevertheless can be used to give a rough, yet still useful, estimate of the host galaxy's luminosity. Following the strategy of Greene & Ho (2004, 2007b), we obtain the host galaxy luminosity by subtracting the AGN contribution, derived from the spectral analysis, from the total Petrosian (galaxy plus AGN) luminosity available from the photometry. In the current application, we use the broad H$\alpha$ luminosity as a surrogate for the 5100 Å continuum luminosity to minimize the uncertainty of measuring the latter, since in some of our objects there may be significant starlight within the 3$''$ aperture of the SDSS spectra (see Greene & Ho 2005b). We extrapolate the flux density at 5100 Å to the central wavelength of the *g* filter (5120 Å) assuming that the underlying power-law continuum has a shape $f_\lambda \propto \lambda^{-1.56}$ (Vanden Berk et al. 2001), adding the small offset to the photometric zeropoint of the *g*-band filter recommended in the SDSS website[8]. In a few sources the host galaxy luminos-ity derived in this manner actually exceeds the total luminosity. This may reflect the inherent scatter introduced by our procedure, or perhaps variability in the AGN. For these cases, we adopt the total luminosity as an upper limit on the host galaxy luminosity.

3. DISCUSSION AND SUMMARY

We have used the Arecibo telescope to conduct the largest modern survey to date for H I emission in active galaxies. The sample consists of 113 $z \lesssim 0.11$ galaxies with type 1 AGNs, selected from an extensive study of SDSS sources for which BH masses can be reliably determined. The new observations were supplemented with an additional 53 type 1 AGNs assembled from the literature, forming a final, comprehensive sample of 154 sources with H I detections or useful upper limits. Among the newly observed galaxies, we detected H I in 66 out of the 101 objects that were not adversely affected by RFI, for an overall detection rate of 65%. The H I masses for the detected sources range from $M_{\rm H I} \approx 10^9$ to $4 \times 10^{10}\ M_\odot$, with an average value of $8.6 \times 10^9\ M_\odot$, while upper limits for the undetected objects generally hover around $M_{\rm H I} \approx 10^{10}\ M_\odot$ (Fig. 5*a*). Adding in the literature sample does not appreciably change these values. The host galaxies of the current sample of type 1 AGNs are therefore quite rich in neutral hydrogen. For reference, recall that our Galaxy has a total H I mass of $5.5 \times 10^9\ M_\odot$ (Hartmann & Burton 1997). Since the H I content of galaxies scales with the stellar luminosity in a manner that depends on morphological type (e.g., Roberts & Haynes 1994), Figure 5*b* examines the H I masses normalized to the *B*-band luminosity of the host galaxy. In the case of the SDSS objects, we converted the host galaxy luminosities in the *g* band (§2.3) to the *B* band assuming an average color of $g - B = -0.45$ mag, appropriate for an Sab galaxy (Fukugita et al. 1995), roughly the average morphological type of our sample. The resulting distribution, ranging from $M_{\rm H I}/L_B \approx 0.02$ to 4.5 with an average value of 0.42, agrees well with the distribution of inactive spiral galaxies of Hubble type Sa to Sb (e.g., Roberts & Haynes 1994). This reinforces the conclusion that the host galaxies of type 1 AGNs possess a normal gas content, at least as far as neutral atomic hydrogen is concerned.

---
[8] http://photo.astro.princeton.edu/#data_model

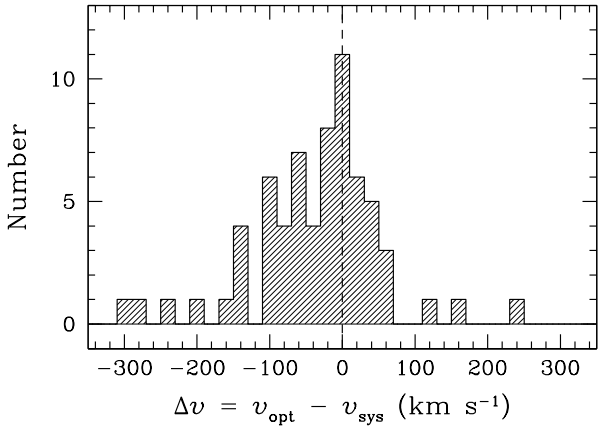

FIG. 6.— Distribution of radial velocity difference as measured in the optical and in H I, $\Delta v = v_{\rm opt} - v_{\rm sys}$. Note the excess of objects toward negative values of $\Delta v$.

The implications of these detection statistics, along with an extensive analysis of the H I and AGN properties assembled here, are presented in our companion paper (Ho et al. 2008).

Figure 6 compares the systemic radial velocity measured from H I with the published optical radial velocity, $v_{\rm opt} = cz$. The velocity difference, $\Delta v = v_{\rm opt} - v_{\rm sys}$, shows a large spread, from $\Delta v \approx -300$ to $+250$ km s$^{-1}$, but there is a noticeable excess at negative velocities. On average, $\langle \Delta v \rangle = -46 \pm 91$ km s$^{-1}$. A similar effect was previously reported by Mirabel & Wilson (1984) and Hutchings et al. (1987); in their samples, the mean offset is $\langle \Delta v \rangle \approx -50$ km s$^{-1}$, essentially identical to our result. Since our sources are relatively bright, type 1 AGNs, the optical radial velocities are predominantly derived from the narrow emission lines. The systemic velocity of the galaxy, on the other hand, is well anchored by the H I measurement. The negative value of $\langle \Delta v \rangle$ therefore implies that on average the ionized gas in the narrow-line region has a general tendency to be mildly outflowing.

The work of L. C. H. was supported by the Carnegie Institution of Washington and by NASA grants HST-GO-10149.02 and HST-AR-10969 from the Space Telescope Science Institute, which is operated by the Association of Universities for Research in Astronomy, Inc., for NASA, under contract NAS5-26555. Support for J. D. and J. E. G. was provided by NASA through Hubble Fellowship grants HF-01183.01-A and HF-01196, respectively, awarded by the Space Telescope Science Institute. We made use of the databases in HyperLeda (http://leda.univ-lyon1.fr/), the Sloan Digital Sky Survey, and the NASA/IPAC Extragalactic Database (http://nedwww.ipac.caltech.edu/), which is operated by the Jet Propulsion Laboratory, California Institute of Technology, under contract with NASA. We thank Minjin Kim for help with preparing the *HST* images and for analyzing the SDSS spectra shown in Table 7, Aaron Barth for sending the *HST* image of PG 0844+349, and C. Motch for making available his published spectra of RX J0602.1+2828 and RX J0608.0+3058. We thank the anonymous referee for helpful suggestions.

## APPENDIX

### LITERATURE DATA

In an effort to assemble a large database of H I parameters for nearby active galaxies with BH mass estimates, we supplemented our new Arecibo observations with a sample drawn from the published literature. While there is no perfect way to accomplish this task, we began by assembling all the H I measurements listed in Hyperleda, which to date contains the most comprehensive and systematic database for this purpose. One limitation of Hyperleda is that it lists H I detections but not upper limits. From this master list we systematically cross-correlated the galaxy names with modern compilations of AGN spectroscopic parameters (e.g., Whittle 1992; Marziani et al. 2003; Boroson & Green 1992), as well as whatever other AGN references known to us, with the goal of finding a matching subset that has reliable measurements of nuclear AGN luminosities and line widths for broad H$\alpha$ or H$\beta$ emission, to be used to calculate BH masses. In total we were able to locate 61 objects, of which 53 are not included in our new survey (the eight overlapping objects are given in Table 3).

Table 5 summarizes the basic properties of the literature sample, as given in Hyperleda. (Note that the isophotal diameters and absolute magnitudes of the literature sample have been corrected for internal extinction, as described in Hyperleda. Because of the generally larger distances and more uncertain morphological types of our Arecibo sample, this correction has not been applied to the latter.) Table 6 gathers all the key properties of the sample, including AGN luminosities and line widths, BH masses, Eddington ratios, H I masses, rotation velocities, dynamical masses, and estimates of host galaxy luminosities; these parameters were derived, to the extent possible, following the same precepts used for the main Arecibo sample. Ho (2007a; see Appendix) discusses some complications encountered in using the inclination angles and rotation velocities given in Hyperleda; this study follows the procedures outlined in that paper. Finally, three of the objects do not have published optical spectroscopic parameters but were observed by SDSS. We reanalyzed their optical spectra and present their emission-line measurements in Table 7.

TABLE 5: Literature Sample

| Source Name | Alternate Name | $T$ | $i$ (°) | $D_{25}^c$ (′) | $z$ | $D_L$ (Mpc) | $M_{B_T^c}$ (mag) | $\sigma_* \pm$ error (km s$^{-1}$) |
|---|---|---|---|---|---|---|---|---|
| (1) | (2) | (3) | (4) | (5) | (6) | (7) | (8) | (9) |
| Akn 564 | UGC 12163 | 6.2 | 47.2 | 0.68 | 0.0247 | 107.6 | −21.19 | ⋯ |
| ESO 140−G043 | Fairall 51 | 3.2 | 71.0 | 2.09 | 0.0142 | 57.8 | −20.23 | ⋯ |
| ESO 362−G018 | MCG−05-13-017 | 0.1 | 56.3 | 1.29 | 0.0126 | 50.1 | −19.75 | ⋯ |
| IC 3528 | SDSS J123455.90+153356.1 | 3.0 | 19.5 | 0.42 | 0.0461 | 199.5 | −21.45 | ⋯ |
| IRAS 04312+4008 | | 2.5 | ⋯ | ⋯ | 0.0207 | 89.5 | ⋯ | ⋯ |
| MCG +05-03-013 | UGC 524 | 3.1 | 26.3 | 1.02 | 0.0359 | 155.6 | −21.91 | ⋯ |
| MCG +08-11-011 | UGC 3374 | 4.0 | 37.2 | 2.75 | 0.0205 | 89.9 | −22.51 | 73.3±52.6 |
| MCG +11-08-054 | UGC 3478 | 3.1 | 77.3 | 1.23 | 0.0128 | 57.5 | −21.15 | ⋯ |
| Mrk 6 | IC 450 | −0.6 | 62.7 | 1.38 | 0.0188 | 86.7 | −20.44 | ⋯ |
| Mrk 10 | UGC 4013 | 3.1 | 67.3 | 1.51 | 0.0293 | 128.2 | −22.69 | 142.9±39.6 |
| Mrk 40 | I Zw 26 | −2.0 | 90.0 | 0.81 | 0.0211 | 93.8 | −19.26 | ⋯ |
| Mrk 79 | UGC 3973 | 3.0 | 33.0 | 1.41 | 0.0222 | 97.3 | −21.75 | 130±20 |
| Mrk 315 | II Zw 187 | −4.9 | 60.9 | 0.72 | 0.0389 | 170.6 | −21.89 | ⋯ |
| Mrk 352 | | −2.0 | 33.9 | 0.68 | 0.0149 | 65.2 | −19.51 | ⋯ |
| Mrk 358 | | 4.0 | 42.3 | 0.71 | 0.0452 | 195.0 | −21.71 | ⋯ |
| Mrk 359 | UGC 1032 | 2.5 | 39.5 | 0.78 | 0.0174 | 73.5 | −20.56 | ⋯ |
| Mrk 493 | UGC 10120 | 3.1 | 45.4 | 0.93 | 0.0313 | 138.0 | −21.16 | ⋯ |
| Mrk 506 | | 1.0 | 47.3 | 0.79 | 0.0430 | 187.1 | −21.71 | ⋯ |
| Mrk 533 | NGC 7674 | 3.8 | 26.7 | 1.12 | 0.0289 | 127.1 | −21.93 | 150.3±33.4 |
| Mrk 541 | | −3.2 | 70.5 | 0.89 | 0.0394 | 169.8 | −20.72 | ⋯ |
| Mrk 543 | NGC 7811, III Zw 127 | 9.9 | 23.7 | 0.41 | 0.0255 | 109.1 | −20.53 | ⋯ |
| Mrk 590 | NGC 863 | 1.0 | 28.5 | 1.23 | 0.0264 | 111.7 | −21.61 | 189±6 |
| Mrk 595 | | 1.0 | 52 | ⋯ | 0.0269 | 114.8 | −20.89 | ⋯ |
| Mrk 704 | | −0.9 | 70 | 0.78 | 0.0292 | 128.8 | −21.14 | ⋯ |
| Mrk 705 | Akn 202, VIII Zw 47 | −2.0 | 40.7 | 0.71 | 0.0292 | 124.2 | −20.95 | ⋯ |
| Mrk 728 | | −0.9 | ⋯ | ⋯ | 0.0357 | 153.5 | −20.82 | ⋯ |
| Mrk 739E | NGC 3758 | 3.8 | 37.4 | 0.62 | 0.0299 | 128.8 | −21.24 | ⋯ |
| Mrk 871 | IC 1198 | 2.8 | 62.4 | 0.71 | 0.0337 | 146.6 | −21.63 | ⋯ |
| Mrk 975 | UGC 774 | 2.9 | 51.5 | 0.98 | 0.0496 | 212.8 | −22.28 | ⋯ |
| Mrk 1018 | UGC 1597 | −1.9 | 54 | 0.91 | 0.0424 | 181.1 | −21.73 | 223.1±19.8 |
| Mrk 1040 | NGC 931 | 3.7 | 81.8 | 2.95 | 0.0167 | 72.4 | −21.17 | 154.0±32.6 |
| Mrk 1044 | | 1.4 | 36.9 | 0.83 | 0.0165 | 68.9 | −19.37 | ⋯ |
| Mrk 1126 | NGC 7450 | −0.9 | 20.6 | 1.62 | 0.0106 | 44.7 | −19.36 | 85.2±25.1 |
| Mrk 1239 | | ⋯ | ⋯ | ⋯ | 0.0199 | 83.6 | ⋯ | 263.4±21.1 |
| Mrk 1243 | NGC 3080 | 1.0 | 18.4 | 0.83 | 0.0354 | 152.1 | −21.25 | ⋯ |
| Mrk 1511 | NGC 5940 | 0.9 | 20.4 | 0.81 | 0.0339 | 147.2 | −21.58 | ⋯ |
| NGC 1566 | | 4.0 | 55.8 | 8.13 | 0.0050 | 17.4 | −21.33 | 109.6±7.8 |
| NGC 3516 | | −2.0 | 39 | 2.04 | 0.0088 | 40.9 | −20.75 | 181±5 |
| NGC 3783 | | 1.3 | 28.0 | 2.29 | 0.00973 | 39.3 | −21.13 | 95±10 |
| NGC 4051 | | 4.0 | 36.0 | 5.13 | 0.00234 | 13.1 | −20.01 | 89±3 |
| NGC 4151 | | 2.1 | 21.0 | 3.80 | 0.00332 | 17.1 | −20.17 | 97±3 |
| NGC 4395 | | 8.8 | 34.5 | 4.07 | 0.00106 | 4.3 | −18.19 | 30 |
| NGC 4593 | Mrk 1330 | 3.0 | 36.0 | 2.63 | 0.00900 | 35.6 | −20.87 | 135±6 |
| NGC 5804 | SDSS J145706.79+494008.4 | 3.1 | 26.5 | 1.17 | 0.0138 | 62.8 | −20.45 | ⋯ |
| NGC 6814 | | 4.0 | 21 | 3.55 | 0.00521 | 22.6 | −21.42 | 111.8±17.5 |
| NGC 7213 | | 0.9 | 30.0 | 3.80 | 0.00584 | 22.8 | −20.89 | 181.7±19.6 |

TABLE 5: LITERATURE SAMPLE—*Continued*

| Source Name | Alternate Name | $T$ | $i$ (°) | $D_{25}^{\rm c}$ (′) | $z$ | $D_L$ (Mpc) | $M_{B_{\rm T}^{\rm c}}$ (mag) | $\sigma_*\pm$error (km s$^{-1}$) |
|---|---|---|---|---|---|---|---|---|
| (1) | (2) | (3) | (4) | (5) | (6) | (7) | (8) | (9) |
| NGC 7214 | | 4.4 | 49.5 | 2.00 | 0.0231 | 95.9 | $-22.08$ | $\cdots$ |
| PG 0007+106 | III Zw 2, Mrk 1501 | 2.9 | $\cdots$ | $\cdots$ | 0.0893 | 383.7 | $-22.15$ | $\cdots$ |
| PG 0050+124 | I Zw 1, Mrk 1502 | $\cdots$ | 44.3 | 0.48 | 0.0611 | 261.8 | $-22.98$ | $\cdots$ |
| PG 1119+120 | Mrk 734 | 3.4 | $\cdots$ | $\cdots$ | 0.0502 | 217.8 | $-21.88$ | $\cdots$ |
| PG 1659+294 | Mrk 504 | 1.6 | 60.2 | 0.55 | 0.0359 | 157.8 | $-20.77$ | $\cdots$ |
| PG 2214+139 | Mrk 304 | $-3.0$ | 35.7 | 0.40 | 0.0658 | 283.1 | $-22.78$ | $\cdots$ |
| UGC 3142 | IRAS 04406+2852 | 0.0 | 39.2 | 1.86 | 0.0217 | 93.8 | $-22.36$ | $\cdots$ |

NOTE.— Data collected from Hyperleda. Col. (1) Source name. Col. (2) Alternate name(s). Col. (3) Morphological type index. Col. (4) Inclination angle. Col. (5) Isophotal diameter at a $B$-band surface brightness of 25 mag arcsec$^{-2}$, corrected for inclination and internal extinction, as described in Bottinelli et al. 1995. Col. (6) Redshift. Col. (7) Luminosity distance, derived assuming a Local Group infall velocity of 208 km s$^{-1}$ toward the Virgo cluster and a Hubble constant of $H_0 = 70$ km s$^{-1}$ Mpc$^{-1}$. The distance to NGC 4395 comes from Thim et al. 2004. Col. (8) Total (host galaxy + AGN) absolute $B$-band magnitude, corrected for Galactic extinction, internal extinction, and $K$-correction. Col. (9) Central stellar velocity dispersion, with additional data from Onken et al. 2004 (Mrk 79 and NGC 3783), Nelson et al. 2004 (Mrk 590, NGC 3516, 4051, 4151, and 4593), and Filippenko & Ho 2003 (NGC 4395).

TABLE 6: Properties of the Host Galaxies

| Source Name | FWHM$_{H\beta}$ (km s$^{-1}$) | Ref. | log $L_{H\beta}$ ($L_\odot$) | Ref. | log $M_{BH}$ ($M_\odot$) | log $L_{bol}$ ($L_{Edd}$) | log $M_{H\,I}$ ($M_\odot$) | $v_m$ (km s$^{-1}$) | Ref. | log $M_{dyn}$ ($M_\odot$) | $M_{B,host}$ (mag) | $M_{B,TF}$ (mag) | Notes |
|---|---|---|---|---|---|---|---|---|---|---|---|---|---|
| (1) | (2) | (3) | (4) | (5) | (6) | (7) | (8) | (9) | (10) | (11) | (12) | (13) | (14) |
| Akn 564 | 1301 | 1 | 40.59 | 1 | 6.00±0.5 | −0.49 | 10.05 | 166.7± 7.8 | 20 | 10.61 | −21.14 | −19.71 | ⋯ |
| ESO 140−G043 | 3077 | 1 | 41.05 | 1 | 7.00±0.5 | −1.08 | 10.01 | 130.3± 5.6 | 20 | 10.61 | −19.88 | −18.91 | ⋯ |
| ESO 362−G018 | 4000 | 2 | 40.99 | 2 | 7.19±0.5 | −1.33 | 9.31 | 125.5± 8.6 | 20 | 10.31 | −19.23 | −18.79 | ⋯ |
| IC 3528 | 1574 | 3 | 41.48 | 3 | 6.42±0.5 | −0.12 | 10.18 | 128.5±21.0 | 20 | 10.44 | −21.19 | −18.87 | 1 |
| IRAS 04312+4008 | 1374 | 4 | 42.02 | 13 | 6.84±0.5 | −0.07 | 9.69 | 101.9± 3.4 | 20 | ⋯ | ⋯ | −18.12 | ⋯ |
| MCG +05-03-013 | 4161 | 6 | 41.03 | 6 | 7.25±0.5 | −1.35 | 10.12 | 263.4±10.5 | 20 | 11.34 | −21.85 | −21.20 | ⋯ |
| MCG +08-11-011 | 6002 | 7 | 41.41 | 14 | 7.78±0.5 | −1.55 | 10.18 | 312.0±34.0 | 20 | 11.68 | −22.43 | −21.75 | ⋯ |
| MCG +11-08-054 | 1461 | 4 | 40.67 | 13 | 6.14±0.5 | −0.56 | 9.83 | 169.9± 6.9 | 20 | 10.61 | −21.09 | −19.78 | ⋯ |
| Mrk 6 | 7375 | 7 | 41.43 | 14 | 7.97±0.5 | −1.72 | 8.35 | 270 | 14 | 11.24 | −19.70 | −21.28 | ⋯ |
| Mrk 10 | 2400 | 8 | 41.38 | 14 | 6.97±0.5 | −0.76 | 10.68 | 255.7±31.9 | 20 | 11.40 | −22.63 | −21.10 | ⋯ |
| Mrk 40 | 2000 | 8 | 40.76 | 14 | 6.46±0.5 | −0.80 | 10.36 | 100 | 14 | 10.18 | −18.75 | −18.05 | ⋯ |
| Mrk 79 | 4852 | 15 | 41.84 | 15 | 7.46±0.11 | −0.85 | 9.87 | 167.2±23.5 | 20 | 10.88 | −21.30 | −19.72 | ⋯ |
| Mrk 315 | 3700 | 7 | 41.02 | 14 | 7.14±0.5 | −1.25 | 10.12 | 89.8±15.1 | 20 | 10.30 | −21.83 | −17.70 | ⋯ |
| Mrk 352 | 3800 | 8 | 41.15 | 14 | 7.24±0.5 | −1.23 | 9.41 | 182.9± 8.8 | 20 | 10.47 | −18.35 | −20.02 | ⋯ |
| Mrk 358 | 2235 | 4 | 40.86 | 14 | 6.62±0.5 | −0.87 | 10.27 | 237.0±12.2 | 20 | 11.19 | −21.66 | −20.86 | ⋯ |
| Mrk 359 | 816 | 4 | 40.40 | 14 | 5.48±0.5 | −0.14 | 8.88 | 38.0± 2.6 | 20 | 9.22 | −20.50 | −14.91 | 2 |
| Mrk 493 | 740 | 6 | 41.13 | 4,6 | 5.81±0.5 | 0.18 | 9.87 | 24.7± 2.0 | 20 | 9.19 | −21.00 | −13.51 | ⋯ |
| Mrk 506 | 5643 | 1 | 41.14 | 1 | 7.58±0.5 | −1.58 | 9.71 | 169.7±16.2 | 20 | 10.93 | −21.61 | −19.77 | ⋯ |
| Mrk 533 | 17053 | 1 | 41.18 | 1 | 8.56±0.5 | −2.53 | 10.45 | 373.5±48.7 | 20 | 11.60 | −21.85 | −22.33 | 3 |
| Mrk 541 | 3300 | 8 | 40.84 | 16 | 6.94±0.5 | −1.21 | 10.15 | 182.6±16.7 | 20 | 11.00 | −20.59 | −20.01 | 4 |
| Mrk 543 | 3828 | 7 | 40.23 | 7 | 6.73±0.5 | −1.54 | 9.71 | 255.2±22.6 | 20 | 10.77 | −20.49 | −21.10 | ⋯ |
| Mrk 590 | 2220 | 15 | 41.94 | 15 | 7.42±0.06 | −0.72 | 10.17 | 372.3±43.0 | 20 | 11.58 | −20.91 | −22.32 | ⋯ |
| Mrk 595 | 2360 | 5 | 41.77 | 14 | 7.17±0.5 | −0.62 | 8.92 | 222 | 14 | ⋯ | −19.74 | −20.64 | 5 |
| Mrk 704 | 5500 | 5 | 41.52 | 14 | 7.77±0.5 | −1.44 | 8.89 | 133 | 14 | 10.55 | −20.74 | −18.98 | 5 |
| Mrk 705 | 2387 | 1 | 41.07 | 1 | 6.79±0.5 | −0.86 | 9.22 | 198.4±13.5 | 20 | 10.84 | −20.78 | −20.28 | ⋯ |
| Mrk 728 | 10275 | 1 | 41.54 | 1 | 8.32±0.5 | −1.97 | 9.50 | 169.2±15.5 | 20 | ⋯ | −20.20 | −19.76 | 6 |
| Mrk 739E | 1615 | 6 | 42.25 | 17 | 7.11±0.5 | −0.13 | 9.51 | 234.9±22.9 | 20 | 10.95 | ⋯ | −20.83 | 7 |
| Mrk 871 | 3688 | 1 | 40.91 | 1 | 7.08±0.5 | −1.29 | 9.98 | 251.9±23.3 | 20 | 11.12 | −21.57 | −21.06 | ⋯ |
| Mrk 975 | 4120 | 1 | 41.27 | 1 | 7.38±0.5 | −1.27 | 9.93 | 223.9±35.6 | 20 | 11.32 | −22.21 | −20.67 | ⋯ |
| Mrk 1018 | 5000 | 9 | 40.33 | 14 | 7.02±0.5 | −1.74 | <9.12 | 155 | 14 | 10.90 | −21.71 | −19.48 | 8,9 |
| Mrk 1040 | 4220 | 4 | 41.00 | 14 | 7.25±0.5 | −1.38 | 10.56 | 208.3±12.8 | 20 | 11.27 | −21.05 | −20.44 | ⋯ |
| Mrk 1044 | 1010 | 6 | 41.01 | 4,6 | 6.01±0.5 | −0.13 | 9.65 | 171.8±18.2 | 20 | 10.53 | −18.47 | −19.81 | ⋯ |
| Mrk 1126 | 2750 | 4 | 40.28 | 18 | 6.47±0.5 | −1.23 | 9.56 | 234.2±26.2 | 20 | 10.90 | −19.21 | −20.82 | ⋯ |
| Mrk 1239 | 1075 | 6 | 41.14 | 14 | 6.14±0.5 | −0.14 | 9.89 | 180 | 14 | ⋯ | ⋯ | −19.96 | 9 |
| Mrk 1243 | 3172 | 4 | 41.01 | 18 | 7.00±0.5 | −1.12 | 10.10 | 226.2±12.1 | 20 | 11.11 | −21.14 | −20.71 | ⋯ |
| Mrk 1511 | 5236 | 1 | 40.79 | 1 | 7.32±0.5 | −1.63 | 9.95 | 226.6±10.8 | 20 | 11.09 | −21.53 | −20.71 | ⋯ |
| NGC 1566 | 2583 | 10 | 40.05 | 10 | 6.29±0.5 | −1.26 | 9.89 | 117.2± 4.1 | 20 | 10.59 | −21.32 | −18.57 | ⋯ |
| NGC 3516 | 3353 | 15 | 40.88 | 15 | 7.37±0.13 | −1.60 | <8.42 | 296 | 21 | 11.16 | −20.61 | −21.58 | 9,10 |
| NGC 3783 | 3093 | 15 | 41.32 | 15 | 7.21±0.07 | −1.05 | 9.57 | 125.7±11.4 | 20 | 10.45 | −20.88 | −18.80 | ⋯ |
| NGC 4051 | 1052 | 15 | 39.81 | 15 | 6.02±0.15 | −1.20 | 9.15 | 209.9±10.3 | 20 | 10.77 | −19.98 | −20.46 | ⋯ |
| NGC 4151 | 4248 | 15 | 40.60 | 23 | 7.40±0.05 | −1.88 | 9.46 | 144.6± 4.3 | 20 | 10.43 | −20.03 | −19.25 | ⋯ |
| NGC 4395 | 1500 | 24 | 37.44 | 24 | 5.30±0.12 | −2.57 | 9.31 | 84.0± 1.6 | 20 | 9.39 | −18.19 | −17.49 | 11 |
| NGC 4593 | 3769 | 15 | 40.53 | 25 | 6.74±0.08 | −1.28 | 9.37 | 274.6±17.2 | 20 | 11.15 | −20.81 | −21.34 | ⋯ |
| NGC 5804 | 2176 | 3 | 40.61 | 3 | 6.23±0.5 | −0.70 | 9.24 | 219.3±41.6 | 20 | 10.85 | −20.34 | −20.60 | 1 |
| NGC 6814 | 4200 | 11 | 40.60 | 11 | 7.02±0.5 | −1.50 | 9.60 | 98.2± 1.4 | 20 | 10.19 | −21.38 | −17.99 | 12 |
| NGC 7213 | 3200 | 2 | 40.78 | 2 | 6.88±0.5 | −1.20 | 9.41 | 378.6±44.9 | 20 | 11.40 | −20.79 | −22.38 | ⋯ |
| NGC 7214 | 4500 | 2 | 41.25 | 2 | 7.44±0.5 | −1.35 | 9.79 | 210.7±17.1 | 20 | 11.23 | −21.99 | −20.48 | ⋯ |
| PG 0007+106 | 5100 | 12 | 42.32 | 19 | 8.15±0.5 | −1.11 | 10.37 | 289.5±64.0 | 20 | ⋯ | −21.08 | −21.51 | 3,13 |
| PG 0050+124 | 1092 | 1 | 41.93 | 1 | 6.59±0.5 | 0.10 | 10.34 | 219.2±17.1 | 20 | 11.08 | −22.83 | −20.60 | ⋯ |
| PG 1119+120 | 2171 | 1 | 41.48 | 1 | 6.94±0.5 | −0.64 | 9.75 | 146.6±21.1 | 20 | ⋯ | −21.71 | −19.30 | 6 |

TABLE 6: Properties of the Host Galaxies—*Continued*

| Source Name (1) | FWHM$_{H\beta}$ (km s$^{-1}$) (2) | Ref. (3) | log $L_{H\beta}$ ($L_\odot$) (4) | Ref. (5) | log $M_{BH}$ ($M_\odot$) (6) | log $L_{bol}$ ($L_{Edd}$) (7) | log $M_{H\,I}$ ($M_\odot$) (8) | $v_m$ (km s$^{-1}$) (9) | Ref. (10) | log $M_{dyn}$ ($M_\odot$) (11) | $M_{B,host}$ (mag) (12) | $M_{B,TF}$ (mag) (13) | Notes (14) |
|---|---|---|---|---|---|---|---|---|---|---|---|---|---|
| PG 1659+294 | 2300 | 8 | 40.95 | 18 | 6.69±0.5 | −0.86 | 9.30 | 116.2± 6.1 | 20 | 10.37 | −20.61 | −18.54 | 6 |
| PG 2214+139 | 5131 | 1 | 42.03 | 1 | 7.99±0.5 | −1.21 | 10.24 | 217.4±18.5 | 20 | 11.03 | −22.55 | −20.58 | ⋯ |
| UGC 3142 | 10489 | 4 | 41.04 | 13 | 8.06±0.5 | −2.15 | 9.87 | 208.9± 9.7 | 20 | 11.18 | −22.32 | −20.45 | ⋯ |

NOTE.— Col. (1) Source name. Col. (2) FWHM of the broad H$\beta$ line. Col. (3) Reference for FWHM$_{H\beta}$. Col. (4) Luminosity of the broad H$\beta$ line; in the case of the reverberation mapped objects, the H$\beta$ luminosity was estimated from the published 5100 Å continuum luminosity, using the $L_{H\beta} - L_{5100\text{Å}}$ correlation (Eq. 2) of Greene & Ho 2005b. Col. (5) Reference for $L_{H\beta}$. Col. (6) Black hole mass derived either from the H$\beta$ method of Greene & Ho 2005b or from reverberation mapping; preference is given to the latter if available. Masses from the latter method come from Peterson et al. 2004, except for NGC 4395 (Peterson et al. 2005), NGC 4151 (Bentz et al. 2006), and NGC 4593 (Denney et al. 2006). For consistency with the virial mass zeropoint adopted by Greene & Ho 2005b, all the reverberation mapped masses were reduced by a factor of 1.8 (see footnote 4 in Greene & Ho 2005b). Col. (7) Eddington ratio, assuming $L_{bol} = 9.8 L_{5100\text{Å}}$ (McLure & Dunlop 2004), with $L_{5100\text{Å}}$ estimated from the $L_{H\beta} - L_{5100\text{Å}}$ correlation of Greene & Ho 2005b. Col. (8) H I mass. Col. (9) Maximum rotational velocity, corrected for inclination. Col. (10) Reference for $v_m$. Col. (11) Dynamical mass of the galaxy, derived from $D_{25}^c$ and $v_m$. Col. (12) $B$-band absolute magnitude of the host galaxy, derived from subtracting the AGN contribution from the total luminosity. Col. (13) $B$-band absolute magnitude of the host galaxy, derived from $v_m$ and the Tully-Fisher relation. Col. (14) Notes: (1) Broad-line parameters based on H$\alpha$ instead of H$\beta$; black hole mass derived using the H$\alpha$ method of Greene & Ho 2005. (2) The $W_{20}$ value listed in Giovanelli & Haynes 1993 gives $V_{max} = 159$ km s$^{-1}$. (3) Broad H$\beta$ component seems very weak and ambiguous; omit from sample. (4) H$\beta$ flux is the average of the values of two epochs, as quoted in Phillips 1978. (5) H I flux from Hutchings 1989. (6) No available value for inclination angle. (7) Galaxy pair; H I line width may be affected by the starburst companion Mrk 739W. (8) $V_{max}$ based on optical rotation curve. (9) H I flux from Mirabel & Wilson 1984. (10) Inclination angle from Mulchaey et al. 1992. (11) The inclination angle given in Hyperleda, 90°, is incorrect. The listed value comes from Ho et al. 1997. (12) The inclination angle given in Hyperleda, 85.6°, is incorrect. The listed value comes from Schulz et al. 1994, but it is very uncertain; it can be as low as $i \approx 0°$. The width of the broad H$\beta$ profile of NGC 6814 is extremely variable, by about a factor of 2; the listed value is the average for "group 1" and "group 2" given in Rosenblatt et al. 1994. The listed broad H$\beta$ flux is also the average of these two groups. (13) No $B$-band magnitude given in Hyperleda. (14) H I line width highly uncertain.

REFERENCES.— (1) Marziani et al. 2003; (2) Winkler 1992; (3) SDSS, this paper; (4) Botte et al. 2004; (5) Stirpe 1990; (6) Véron-Cetty et al. 2001; (7) Wang et al. 2006; (8) Osterbrock 1977; (9) Osterbrock 1981; (10) Kriss et al. 1991; (11) Rosenblatt et al. 1994; (12) Boroson & Green 1992; (13) Greene & Ho 2005, converted from $\lambda L_\lambda$(5100 Å) luminosity given in Botte et al. 2004; (14) Whittle 1992; (15) Peterson et al. 2004; (16) Phillips 1978; (17) Netzer et al. 1987; (18) Dahari & De Robertis 1988; (19) L. C. Ho 2007, in preparation; (20) Hyperleda (http://leda.univ-lyon1.fr/); (21) Mulchaey et al. 1992; (22) Condon et al. 1985; (23) Bentz et al. 2006; (24) Filippenko & Ho 2003; (25) Denney et al. 2006.

TABLE 7: Emission-line Properties of New SDSS Objects

| Source Name | Narrow Lines | | | | | | Broad Lines | | | | Type |
|---|---|---|---|---|---|---|---|---|---|---|---|
| | H$\beta$ $\lambda$4861 | [O III] $\lambda$5007 | [O I] $\lambda$6300 | H$\alpha$ $\lambda$6563 | [N II] $\lambda$6583 | [S II] $\lambda\lambda$6716, 6731 | H$\beta$ $\lambda$4861 | FWHM (km s$^{-1}$) | H$\alpha$ $\lambda$6563 | FWHM (km s$^{-1}$) | |
| SDSS J120257.81+045045.0 | 4.68 | 48.1 | 3.81 | 20.5 | 31.5 | 18.9 | 17.9 | 3829 | 151 | 2066 | S1.5 |
| SDSS J123455.90+153356.1 | 33.0 | 158 | 6.49 | 161 | 111 | 55.6 | 118 | 1891 | 640 | 1574 | S1.2 |
| SDSS J145706.79+494008.4 | 58.8 | 36.0 | 13.4 | 199 | 131 | 90.1 | 367 | 3328 | 873 | 2176 | S1.0 |

NOTE.—Line fluxes in units of $10^{-16}$ ergs s$^{-1}$ cm$^{-2}$, corrected for Galactic extinction, but not for internal extinction. The Seyfert type given in the last column is based on the observed flux ratio of [O III] $\lambda$5007/H$\beta$(total), following the criteria recommended by Whittle 1992.